\documentclass[a4paper,twoside,prd,nofootinbib,preprintnumbers,twocolumn]{revtex4-2}
\usepackage{amssymb,amsmath,bm,natbib}
\usepackage{color}
\usepackage{slashed}
\usepackage{graphics}
\usepackage{graphicx}
\usepackage[utf8]{inputenc}
\usepackage{tabularx}
\usepackage[caption=false]{subfig}
\usepackage{hyperref}
\usepackage{url}
\usepackage{dsfont}
\usepackage{float} % pour le placement des images
\usepackage{cancel}
\usepackage{ulem}

\newcommand{\TeV}{\,\text{TeV}}

\begin{document}
	\preprint{PSI-PR-22-27, ZU-TH 42/22}
\title{Consistency and Interpretation of the LHC (Di-)Di-Jet Excesses}

\author{Andreas Crivellin}
\email{andreas.crivellin@cern.ch}
\affiliation{Physik-Institut, Universit\"at Z\"urich, Winterthurerstrasse 190, CH--8057 Z\"urich, Switzerland}
\affiliation{Paul Scherrer Institut, CH--5232 Villigen PSI, Switzerland}
	
\author{Claudio Andrea Manzari}
\email{claudioandrea.manzari@physik.uzh.ch}
\affiliation{Physik-Institut, Universit\"at Z\"urich, Winterthurerstrasse 190, CH--8057 Z\"urich, Switzerland}
\affiliation{Paul Scherrer Institut, CH--5232 Villigen PSI, Switzerland}

\author{Bruce Mellado}
\email{bmellado@mail.cern.ch}
\affiliation{School of Physics and Institute for Collider Particle Physics, University of the Witwatersrand, Johannesburg, Wits 2050, South Africa}
\affiliation{iThemba LABS, National Research Foundation, PO Box 722, Somerset West 7129, South Africa}

\author{Salah-Eddine Dahbi}
\email{salah-eddine.dahbi@cern.ch}
\affiliation{School of Physics and Institute for Collider Particle Physics, University of the Witwatersrand, Johannesburg, Wits 2050, South Africa}

%\author{Abhaya Kumar Swain}
%\email{abhaya.kumar.swain@cern.ch}
%\affiliation{School of Physics and Institute for Collider Particle Physics, University of the Witwatersrand, Johannesburg, Wits 2050, South Africa}

\begin{abstract}
ATLAS observed a limit for {the cross section of di-jets resonances, which is weaker than expected for a} mass slightly below $\approx$1\TeV. In addition, CMS reported hints for the (non-resonant) pair production of di-jet resonances $X$ via a particle $Y$ at a very similar mass range with a local (global) significance of 3.6\,$\sigma$ (2.5\,$\sigma$) at $m_X\approx950\,$GeV. In this article we show that using the preferred range for $m_X$ from the ATLAS analysis, one can reinterpret the CMS analysis of di-di-jets in terms of a resonant search with $Y\to XX$, with a significantly reduced look-elsewhere effect, finding an excess for $m_Y\!\approx\!3.6$\TeV with a significance of $4.0\,\sigma$ ($3.2\,\sigma$) locally (globally). We present two possible UV completions capable of explaining the (di-)di-jet excesses, one containing two scalar di-quarks, the other one involving heavy gluons based on an $SU(3)_1\!\times\! SU(3)_2\!\times\! SU(3)_3$ gauge symmetry, spontaneously broken to $SU(3)$ color. In the latter case, non-perturbative couplings are required, pointing towards a composite or extra-dimensional framework. In fact, using 5D-AdS space-time, one obtains the correct mass ratio for $m_X/m_Y$, assuming the $X$ is the lowest lying resonance, and predicts a third (di-)di-jet resonance with a mass around $\approx2.2$\TeV.
\end{abstract}
\maketitle

\section{Introduction}
\label{intro}
Since the discovery of the Higgs boson in 2012~\cite{Aad:2012tfa,Chatrchyan:2012ufa} the main focus of the LHC has been on the discovery of new particles and new interactions beyond the ones included in the Standard Model (SM) of particle physics. While intriguing indirect signs emerged (see e.g.~Refs.~\cite{Crivellin:2022qcj,Crivellin:2021sff,Fischer:2021sqw} for recent reviews of lepton flavour universality violation), no new resonance has been discovered yet. However, recently the number of hints for new physics (NP) in direct LHC searches increased. In particular, ATLAS~\cite{ATLAS:2018qto} observed a weaker limit than expected in resonant di-jet searches\footnote{The analogous CMS di-jet search~\cite{CMS:2018mgb} does not display an excess in the same region. However, the sensitivity is significantly lower, such that the signal suggested by the ATLAS analysis is not excluded.} in a mass region slightly below 1\TeV, while CMS~\cite{CMS:2022usq} found hints for the (non-resonant) pair production of di-jet resonances with a mass of $\approx 950$\,GeV (see Appendix) with a local (global) significance of 3.6\,$\sigma$ (2.5\,$\sigma$) when integrating over the di-di-jet mass.

While the ATLAS analysis by itself does not constitute a significant hint for beyond the SM physics once the look-elsewhere effect (LEE) is taken into account, the compatibility of the suggested di-jet mass with the one of the (non-resonant) CMS di-di-jet analysis is very good. This agreement suggests that both excesses might be due to the same new particle $X$, once directly (resonantly) produced in proton-proton collisions ($pp\!\to\! X\!\to\! jj$), once pair produced via a new state $Y$ ($pp\!\to\! Y^{(*)}\!\to\! XX\!\to\! (jj)(jj)$). While the CMS collaboration in their analysis interprets the di-di-jet excess as the non-resonant production of $XX$ (with $m_X\approx 950\,$GeV) via a heavy new particle $Y$, with $m_Y\approx 8$\,TeV, resulting in a local (global) significance of $3.9\sigma$ ($1.6\sigma$)~\cite{CMS:2022usq}, it is also possible that the two $X$ particles are resonantly produced from the decay of an on-shell $Y$ particle. In fact, the CMS results suggest 3\,TeV$\lessapprox \!m_Y\!\lessapprox$4\,TeV (see Appendix) for such a resonant scenario, once $m_X$ is assumed to be within the preferred range of the ATLAS di-jet analysis. 

\begin{figure}
	\centering 
	\includegraphics[width=0.4\textwidth]{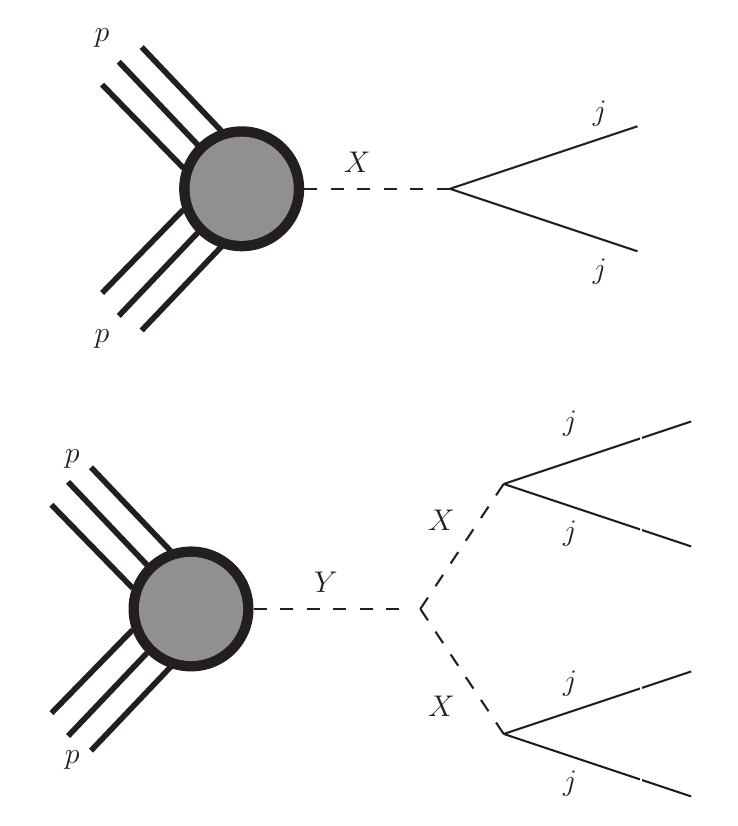}
	\caption{Feynman diagrams showing the resonant production of di-jets via the particle $X$ (upper panel) and di-di-jets via the decay chain $Y\!\to\!XX\!\to\!(jj)(jj)$ (lower panel). Note that $X$ and $Y$ could be scalar or vector bosons in our setup. As we show in the main text, $m_X\!\approx\!950\,$GeV and $m_Y\!\approx\!3.6\,$TeV is preferred by the combination of the ATLAS and CMS analyses.}
	\label{fig:Feynman}
\end{figure}

In order to evaluate this option more quantitatively, a (at least simplified) model is necessary such that the experimental resolution and acceptance can be simulated. We will do this in Sec.~\ref{sec:di-di-jet} using a simplified model with new vector bosons in order to derive the significance resulting from the CMS analysis for such a scenario with an on-shell $Y$ resonance decaying to two $X$ particles, as illustrated in Fig~\ref{fig:Feynman}. Next, we will examine possible UV completions that can provide a common explanation of the (di-)di-jet excesses. As we will discuss in Sec.~\ref{sec:model}, two scalar di-quarks or new massive gluons seem to be the most plausible candidates. Concerning the latter, we will consider a specific example based on an $SU(3)_1\times SU(3)_2\times SU(3)_3$ gauge group, broken down to $SU(3)$ color via two bi-triplets. We then conclude and present an outlook in Sec.~\ref{sec:conclusions}.

\section{(Di-)Di-Jets}
\label{sec:di-di-jet}

As outlined in the introduction, the preferred value for the di-jet invariant mass of ATLAS and CMS analyses strongly suggest that both signals are due to the same particle $X$, i.e.~that $pp\!\to\! X\to jj$ and $pp\!\to\! Y\!\to\! XX\!\to\! (jj)(jj)$ account for the di-jet and the di-di-jet excess, respectively (see Fig.~\ref{fig:Feynman}). In this section we consider this setup within a simplified model with a vector boson $Y$ decaying into two vector bosons $X$.\footnote{In the next section we will consider models that could provide a common explanation of the (di-)di-jet excesses. There we will also consider a model with scalars. We did not explicitly simulate this setup, however the differences compared to the case with gauge bosons is expected to be small as the decay kinematics are very similar.} We will assume that the vectors have a $Y-X-X$ coupling, depending on the momenta in the same way as the SM $Z-W-W$ coupling, with $m_Y>m_X$ and Br$[Y\to XX]=$100\%. In addition to this triple gauge boson interaction, only $X$ and $Y$ couplings to SM quarks, which we assume to be flavour universal, are relevant.

First of all, we fix $900 \,{\rm GeV} \lessapprox m_{X}\lessapprox1050{\rm GeV}$ from the invariant mass preferred by the di-jet analysis of ATLAS~\cite{ATLAS:2018qto} which is based on 29.3\,fb$^{-1}$ integrated luminosity at 13\,TeV.\footnote{See e.g.~Refs.~\cite{Han:2009ya,Gogoladze:2010xd,Giudice:2011ak} for theory accounts of (di-)di-jet searches.} Note that we do not include the significance of the ATLAS measurement in our fit but rather use it to confine ourselves to this range, which reduces the LEE with respect to the di-jet invariant mass. We then employ $m_{X}=950$\,GeV, which corresponds to the best value obtained in the non-resonant analysis by CMS. As such, we move on to the di-di-jet mass $m_Y$ for which the CMS search for pairs of jets was performed with 139\,fb$^{-1}$ integrated luminosity at 13\,TeV center of mass energy~\cite{CMS:2022usq}. In this analysis, {CMS selected} four high transverse momentum jets, including both the cases of resonant $pp\!\to\! Y\!\to\! XX\!\to\! 4j$ and non-resonant $pp\!\to\! XX\!\to\! (jj)(jj)$ production. The observable 
\begin{equation}
    \alpha = \frac{m_1+m_2}{2\cdot m_{4j}}\,,
\end{equation}
is defined, where $m_1$ and $m_2$ are the di-jet invariant masses and $m_{4j}$ is the invariant mass of the four-jet system. The search is then performed in bins of $\alpha$, and in the non-resonant case an excess at $m_Y\approx8.5\,$TeV with a local (global) significance of 3.9\,$\sigma$ (1.6\,$\sigma$) is reported. However, also a resonant-like excess in the four-jet invariant mass spectrum around 3-4\,TeV, i.e.~for $\alpha=0.27, 0.29, 0.31$ with $m_X\approx 950\,$GeV, is visible. The cross-section of this four-jet excess can naively be estimated to be of the order of $O$(fb). 

{The dominant background for di-jet resonance searches in proton-proton collisions is QCD production of multi-jets. For both ATLAS and CMS, Monte-Carlo simulations of this background are used for signal optimization and to provide an approximate comparisons with the observed data. The generation of multi-jets background is realized by simulating the leading order QCD $2\!\to\! 2$ processes of jet production, including extra jets from QCD initial and final state radiation in the parton shower level. In order to avoid the miss-modeling of the multi-jets background, which is closely connected to the detector identification and isolation requirements, the final normalisation and shape of this background is estimated from data by ATLAS and CMS using a data-driven method, described and detailed in Refs.~\cite{CMS:2019gwf,ATLAS:2019fgd}.}

In order to evaluate this possibility of a resonant production of $X(950)$ more quantitatively, we use our simplified model to simulate $pp\!\to\! Y\!\to\!XX\to (jj)(jj)$ events using \texttt{MadGraph5$\_$aMC@NLO 2.6.7} with leading order (LO) accuracy in QCD~\cite{Alwall:2014hca}. The parton showering and hadronization are simulated with \texttt{PYTHIA 8.2}~\cite{Sjostrand:2014zea} using the NNPDF2.3 LO parton distribution function set~\cite{Ball:2012cx}. The events were processed with \texttt{Delphes~3}~\cite{deFavereau:2013fsa}, which provides an approximate fast simulation of CMS detector. Jets were reconstructed using the anti-$kt$ algorithm~\cite{Cacciari:2008gp} with the radius parameter $R=0.4$, as implemented in \texttt{FastJet}~3.2.2~\cite{Cacciari:2011ma}. Jets with $p_{\textrm{T}} > 80$\,GeV and $|\eta| < 2.5$ are considered. Reconstructed jets overlapping with photons, electrons or muons in a cone of size $R=0.4$ are then removed. The four jets with the highest $p_{\textrm{T}}$ are considered as the leading jets. Then the most probably di-jet pairs combination are created by minimizing the $\eta-\phi$ space separations of the jets in each events: 
\begin{equation}
    \Delta R = |(\Delta R_{1}- 0.8)| + |(\Delta R_{2}- 0.8)|\,,
\end{equation}
where $\Delta R_{1}$ and $\Delta R_{2}$ are the $\eta-\phi$\footnote{The distance $\Delta R$ between two jets in the $\eta-\phi$ space is defined as $\Delta R=\sqrt{(\Delta\eta_{jj})^2 + (\Delta\phi_{jj})^2}$.} space separations between the two jets within the respective systems. The offset of 0.8 is chosen to avoid the pairings with overlapped jets. In addition, we require the $\Delta R_{i; i=1, 2}$ to be less than 2, in order to reject contribution from hard jets produced by QCD processes. While the pseudo-rapidity separation $\Delta\eta_{jj}$ between the two jets of each di-jet systems is required to be below 1.1, to remove contribution of backgrounds from QCD \textrm{t}-channel. In the end, we required the asymmetry in the di-jet mass between the di-jet systems to be small ($\frac{|m_1 - m_2|}{m_1 + m_2} < 0.1$) which essentially select the di-jets of equal mass taking into account the energy resolution. This, in turn, is the property of a pair of equal mass resonances, which is unlike to QCD jets that constitute the SM background. 

\begin{figure*}
	\centering 
		\includegraphics[width=0.355\textwidth]{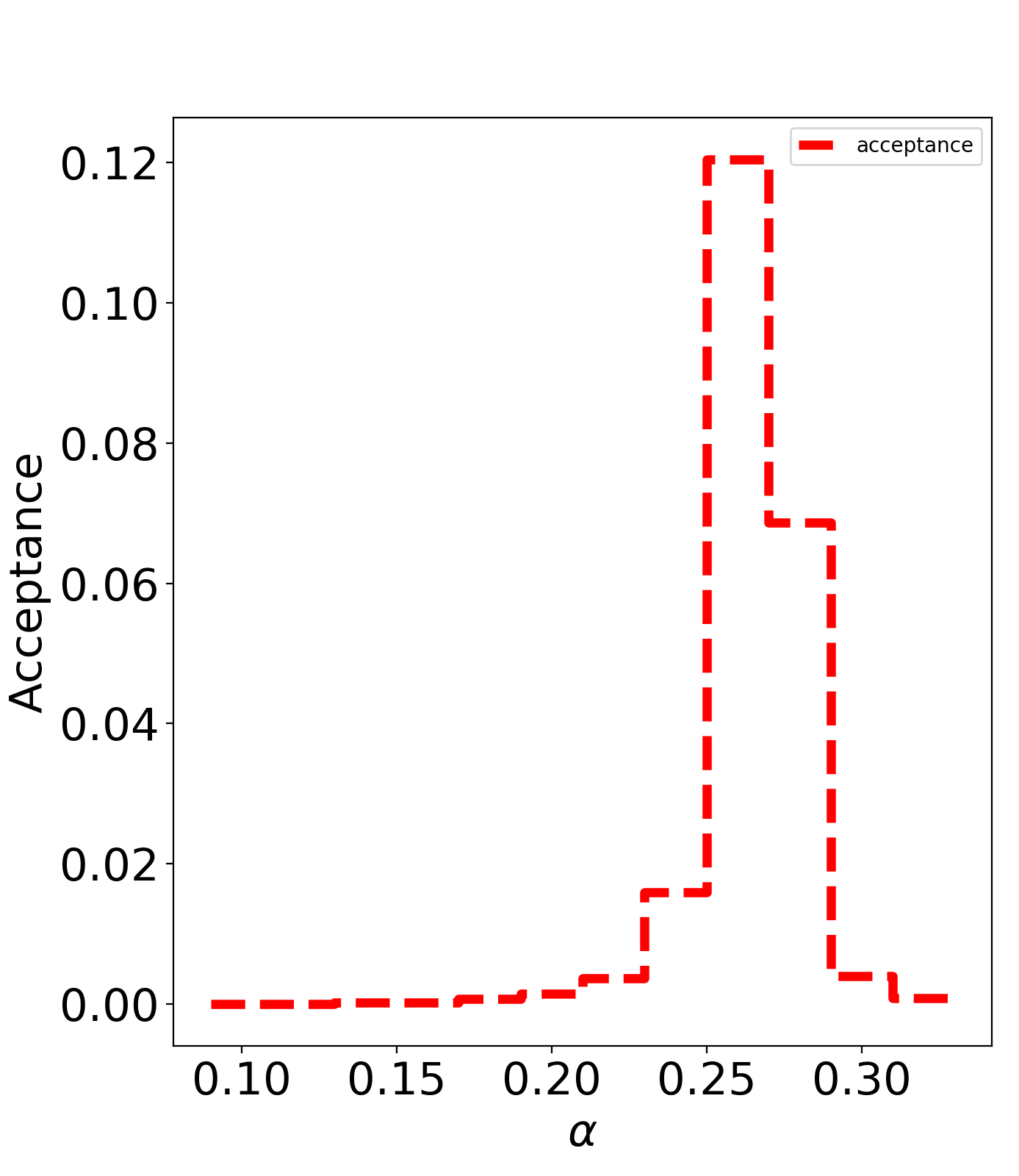}~~~~~~
	\includegraphics[width=0.65\textwidth]{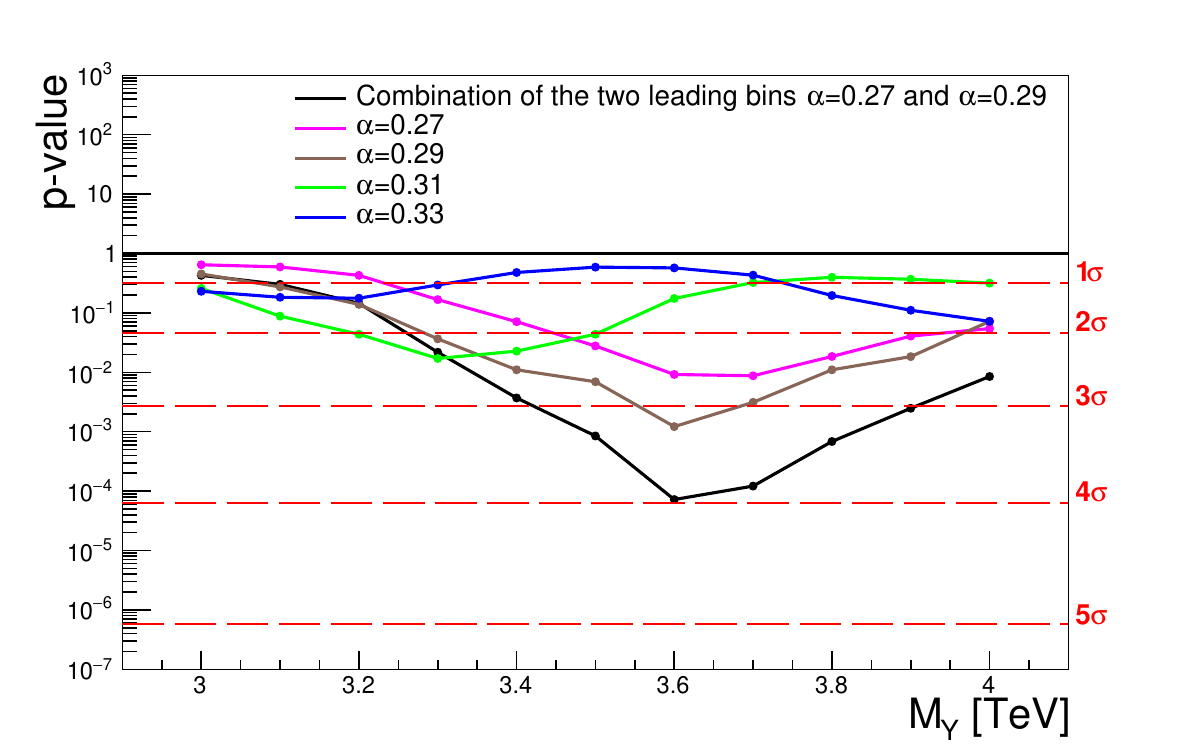}
	\caption{{Left: Acceptance obtained from our simulation of $pp\to Y\to XX\to 4j$ for $m_X=1$~TeV and $m_Y=3.5$~TeV. \newline Right: p-value as a function of $m_Y$, obtained by combining the two leading bins in $\alpha$, i.e.~$\alpha=0.27$ and $\alpha=0.29$.} }
	\label{p-value}
\end{figure*}

The most significant signal in the CMS analysis is found in the bins with the central values $\alpha=0.27$ and $\alpha=0.29$. We therefore evaluated the acceptance and the resolution by simulating the process $pp\!\to\! Y\!\to\!XX\to (jj)(jj)$. The results for $m_Y=3.5\,$TeV and $m_X=1\,$TeV is shown in left panel of Fig.~\ref{p-value}. Because the number of NP events in the two bins is correlated, as given by the acceptance, we can write the $p$-value\footnote{See e.g.~Ref.~\cite{Lyons2020} for the statistical combination of the results from two or more measurements.} of the weigthed average of the two dominant bins as
\begin{equation}
   p=2\times[1 - \Phi\left(\frac{\sum_{i=1}^2 w_i S_i}{\sqrt{\sum_{i=1}^2 w_i^2}}\right)], 
\end{equation}
where $S_i$ is the significance for the $i^{\rm th}$ bin (given in standard deviations) and the weight $w_i$ is equal to the acceptance of each bin, while $\Phi(x)=\frac{1}{\sqrt{2 \pi}}\int_{-\infty}^x  e^{-x^{\prime 2} / 2} d x^{\prime}$ denotes the standard normal cumulative distribution function. From the right panel of Fig.~\ref{p-value}, we can see that the best agreement with data is found for $m_{Y}\approx 3.6\,$TeV, with a total cross-section for $pp\to Y\to XX\to jjjj$ of $\approx5\,$fb. The corresponding local (global) significance is $4\,\sigma$ ($3.2\,\sigma$). Note that the global significance of our resonant excess is higher than the non-resonant effect of CMS mainly due to the smaller LEE as we fixed the range of the di-jet mass a priori with the help of the ATLAS data. The LEE effect evaluated here includes the range $m_Y$ used in the search. 

\section{Interpretation}
\label{sec:model}

\subsection{Vector Bosons Based on $SU(3)^3$}

A model with new vector bosons seems a natural possibility for providing a common explanation of the di-di-jet excesses as such states can have sizable couplings to valence quarks without breaking $SU(2)_L$ (similar to the SM gauge bosons), and in fact already coupling of the order $10^{-1}$ turn out to be sufficient to obtain suitable cross sections. Since self-interactions are required to give rise to $Y\to XX$, this suggests that the new heavy vector bosons originate from a non-abelian gauge group. Furthermore, if one wants that all new vectors to couple to quarks, they must have the same quantum numbers as the $SU(2)_L$ or $SU(3)_c$ gauge bosons of the SM, since otherwise the operators $V_\mu^a \bar q\gamma^\mu T^a q$, where $V_\mu^a$ are the new vector bosons and $T^a$ the corresponding generators, would not be invariant under the SM gauge group. In addition, in order to avoid couplings to leptons, which are strongly constrained from di-lepton searches~\cite{ATLAS:2020yat,CMS:2021ctt}, as well as bounds from electroweak precision observables~\cite{ALEPH:2005ab,ALEPH:2013dgf}, we will opt for a gauge group based on, and related to, $SU(3)_c$. 

Models with such additional heavy colored states based on an extended group for the strong interactions, whose spontaneous symmetry breaking reduces it to its diagonal subgroup, then identified with $SU(3)_c$, were proposed and studied in Refs.~\cite{Hill:1991at,Frampton:1987dn,Chivukula:1996yr,Martynov:2009en,Frampton:2009rk,Chivukula:2013kw}. Furthermore, such a setup emerges in the context of extra space-time dimensions where Kaluza-Klein excitations of gluons exist~\cite{Han:1998sg,Dicus:2000hm,Davoudiasl:2000wi,Lillie:2007yh} and a similar picture arises in composite/technicolor models~\cite{Dimopoulos:1979sp,Farhi:1980xs,Hill:1992ev}.

As we need heavy resonances with (at least) two different masses, we consider the gauge group 
\begin{equation}
SU(3)_1\times SU(3)_2\times SU(3)_3\times SU(2)_L\times U(1)_Y\,,
\end{equation}
broken down to the SM one $SU(3)_c\times SU(2)_L\times U(1)_Y$ via two bi-doublet charged under two non-identical $SU(3)$ groups, each. Here we use
\begin{align}
    \begin{tabular}{c|c c c c c}
         & $SU(3)_1$ & $SU(3)_2$ & $SU(3)_3$ \\
         \hline
                $\Omega_{12}$ & 3 & $\bar 3$ & 1 \\
             $\Omega_{23}$ & 1 & 3 & $\bar 3$ 
    \end{tabular}\qquad {\rm with}\nonumber\\
\left\langle {{\Omega _{12}}} \right\rangle  = {v_{12}}\left( {\begin{array}{*{20}{c}}
1&0&0\\
0&1&0\\
0&0&1
\end{array}} \right),\;\left\langle {{\Omega _{23}}} \right\rangle  = {v_{23}}\left( {\begin{array}{*{20}{c}}
1&0&0\\
0&1&0\\
0&0&1
\end{array}} \right),
\end{align}
which constitutes a choice of basis, i.e.~any other combination ($12,\,13$ or $13,\,23$) would lead to the same physical results. The spontaneous symmetry breaking via the vacuum expectation values $v_{12}$ and $v_{23}$ leads to the following mass matrix for the $SU(3)$ gauge fields $G^{\mu a}_i$ ($i=1,2,3$) in the interaction basis
\begin{widetext}
\begin{equation}
L_M^G =  \frac{1}{2}{\left( {\begin{array}{*{20}{c}}
{G_1^{\mu a} }\\
{G_2^{\mu a} }\\
{G_3^{\mu a} }
\end{array}} \right)^T}\left( {\begin{array}{*{20}{c}}
{v_{12}^2g_1^2}&{v_{12}^2g_1^{}g_2^{}}&0\\
{v_{12}^2g_1^{}g_2^{}}&{\left(v_{12}^2 + v_{23}^2\right)g_2^2}&{v_{23}^2{g_2}{g_3}}\\
0&{v_{23}^2{g_2}{g_3}}&{v_{23}^2g_3^2}
\end{array}} \right)\left( {\begin{array}{*{20}{c}}
{G_1^{\mu a} }\\
{G_2^{\mu a} }\\
{G_3^{\mu a} }
\end{array}} \right)\,,
\end{equation}
\end{widetext}
where each block corresponds to $a=1,...,8$ gauge bosons with the corresponding generators $T^a$ and equal masses. 

We can now diagonalize this mass matrix to obtain the mass eigenstates, $g_1^{a\mu}$, $g_2^{a\mu}$ and $g_3^{a\mu}$ and identify the state with the zero eigenvalue $g_1^{a\mu}$ with the SM gluons and the corresponding coupling with the strong coupling constant $g_s$. The mass of $g_2^{a\mu}$ ($g_3^{a\mu}$) should correspond to the $X$ ($Y$) resonance, i.e.~$950\,$GeV ($3.6\,$TeV). We can furthermore determine the couplings of $g_2^{a\mu}$ and $g_3^{a\mu}$ by demanding that the correct signal strengths are obtained. Since ATLAS finds a preferred value of $g_q\approx0.07$ (in their conventions where quarks couple only to the axial-vector current) for the $X$ resonance, and in our model we have 8 $g_2^{a\mu}$ fields which couples each vectorially and flavour universal to quarks, we find that the production cross section is 4 times larger (for equal couplings) resulting in 
$g^\prime\approx0.035$, where $g^\prime$ ($g^{\prime\prime}$) is the (effective) coupling of $g_2^{a\mu}$ ($g_3^{a\mu}$) to SM quarks. The preferred value for the di-di-jet cross section obtained in the last section is $\approx5fb$. From this we find $g^{\prime\prime}\approx 0.07/\sqrt{{\rm Br}[g^{a\mu}_3\to g^{a\mu}_2g^{a\mu}_2]},$ by using the total production cross section for a sequential SM $Z^\prime$ of this mass ($20\,$fb~\cite{ATLAS:2019vcr}) and taking into account the $Z^\prime$ branching ratio and the PDF scaling, using the PDF of Ref.~\cite{Ball:2013hta} implemented in \texttt{ManeParse}~\cite{Clark:2016jgm}, in order to rescale the cross section to the one of our model.

We can now attempt to solve this system of equations if one specifies under which $SU(3)_i$ gauge factors the SM quarks transform as a triplet. There are seven possibilities for such charge assignments ($SU(3)_1$, $SU(3)_2$, $SU(3)_3$, $SU(3)_1\!\vee\! SU(3)_2$, $SU(3)_1\!\vee\! SU(3)_3$, $SU(3)_1\!\vee\! SU(3)_3$ and $SU(3)_1\!\vee\! SU(2)_1\!\vee\! SU(3)_3$) among which only the option that the SM quarks are $SU(3)_1$ triplets, but uncharged under both other $SU(3)$ gauge factors, provides a solution. In fact, we find $g_1\approx 1$, $g_2\approx 10$, $g_3\approx 15$ which is clearly in the non-perturbaitve regime. Therefore, these values should not be taken at face value, but rather only show that the system of equations has a solution. These large values for the couplings $g_2$ and $g_3$ can be traced back to the smallness of the $g_2^{a\mu}$ and $g_3^{a\mu}$ couplings to SM quarks which requires small mixing among the colored gauge bosons. Nonetheless, as the decay width to SM fermions is small and the right masses and couplings can be obtained, this suggests that the gauge group $SU(3)_1\times SU(3)_2\times SU(3)_3$, broken to $SU(3)_c$ via the described breaking, can in fact explain the (di-)di-jet excesses. Furthermore, the sizable couplings $g_2$ and $g_3$ point towards an extra-dimensional or composite realization of this setup.

\subsection{Scalar Di-Quarks}

Alternatively to the vector-boson model proposed above, one could try to find a perturbative explanation of the (di-)di-jet excesses using scalar bosons. Because the suggested cross sections are too large to originate from a scalar produced via gluon fusion (with perturbative couplings)~\cite{LHCHiggsCrossSectionWorkingGroup:2016ypw}, relevant couplings to valence quarks are needed. Since $SU(3)_c$ singlet scalars can only interact with quarks in the same way as the SM Higgs boson, the couplings are naturally related to the respective Yukawa couplings, rendering them tiny for valence quarks, thus resulting in too small cross sections. 

However, $SU(3)_c$ triplet or sextuplet (symmetric $3\times 3$) scalars can couple to quarks of the same $SU(2)_L$ representation such that their couplings are unrelated to EW symmetry breaking and therefore also unrelated to quark Yukawa couplings. Searches for such di-quarks via di-jet and di-di-jet signatures were proposed in Refs.~\cite{Cakir:2005iw,Mohapatra:2007af,Chen:2008hh,Kilic:2008pm,Berger:2010fy,Bai:2011mr,Dobrescu:2018psr,Pascual-Dias:2020hxo}. 

The choice of quantum numbers for di-quarks is restricted to five possibilities
\begin{equation}
    \begin{tabular}{c|c c c}
         & $SU(3)_c$ & $SU(2)_L$ & $U(1)_Y$\\
         \hline
        $\Phi_{u}$ & $\bar 6$ & 1 & -4/3\\
        $\Phi_{d}$ & $\bar 6$ & 1 & 2/3\\
        $\Phi_{q}^{(1)}$ & 3 & 1 & -1/3\\
        $\Phi_{q}^{(3)}$ & $\bar 6$ & 3 & -1/3\\
        $\Phi_{ud}$ & $\bar 6$ & 1 & 1 \\
    \end{tabular}
    \label{tab:Scalar}
\end{equation}
if we restrict ourselves to the cases which allow couplings symmetric in flavour space. Note that anti-symmetric couplings would in general cause problems with $\Delta F=2$ processes. In this case we have the coupling to fermions
\begin{align}
    \mathcal{L}_{\rm int} =& \lambda_u \bar{u}_R^c\Phi_u u_R + \lambda_d \bar{d}_R^c\Phi_d d_R + \lambda_q^{(3)} \Phi_q^{(3)I}\,\bar{q}_L^c i\sigma_2\,\tau^I q_L\nonumber\\
    &+ \lambda_q^{(1)}\epsilon \Phi_q^{(1)}\,\bar{q}_L^c i\sigma_2\, q_L+ \lambda_{ud} \bar{u}_R^c\Phi_{ud} d_R + {\rm h.c.}\,,
\end{align}
where $u_R,\, d_R$ and $q_L$ are the SM right-handed $SU(2)_L$ singlet quarks and left-handed $SU(2)_L$ doublet quarks, respectively. $\epsilon$ is the totally anti-symmetric tensor in three dimensions which contracts the implicit color indices of the color triplets and we suppressed flavour indices. 

We can thus attempt to construct a scenario which has the potential to reproduce the experimental signals. Assuming that $\Phi_u$ is the $3.6\,$TeV resonance decaying into two $\Phi_d$ scalars with a mass of $950\,$~GeV each, the interaction term $A\,\epsilon \Phi_u\Phi_d\Phi_d$ is needed where the first (second) $\epsilon$ contracts the first (second) $SU(3)_c$ index of the symmetric $3\times3$ representations. Since both scalars couple to right-handed $SU(2)_L$ singlet quarks, we can assume that they have flavour diagonal couplings, both in the interaction and in the mass basis. However, as the couplings to first generation quarks are constrained by neutron-anti-neutron oscillations~\cite{Baldo-Ceolin:1994hzw}, one has to assume that the couplings to second generation quarks are dominant (at least for either $\Phi_u$ or $\Phi_d$). These couplings are then determined by requiring that the correct signal strengths are obtained.\footnote{Note that neither the CMS nor the ATLAS analysis is sensitive to the electric charge of the vector because the jet charge is not measured. In addition, the differences in efficiencies between scalar and vector resonances are expected to be small for the analyses under investigation.} The ATLAS di-jet analysis gives $\vert \lambda_{d(s)} \vert \simeq 0.05(0.2)$. In addition, assuming ${\rm Br}(\Phi_u \to \Phi_d\Phi_d)\simeq 100\%$, which is natural for $A=O({\rm TeV})$), we find $\vert \lambda_{u(c)} \vert \simeq 0.02(1.1)$ from the di-di-jet cross section of $\approx 5 \,$fb.

In principle, also the option that $\Phi_d$ is the $3.6\,$TeV resonance and $\Phi_q^{(3)}$ or $\Phi_q^{(1)}$ the $950\,$GeV one is possible. In this case, it has to be assumed that the couplings to quarks are universal, such that the CKM rotation between the interaction and the mass eigenbasis does not generate flavour changing couplings that would contribute to $\Delta F=2$ processes.

\section{Conclusions and Outlook}
\label{sec:conclusions}

In this article we pointed out that the ATLAS di-jet excess with a mass slightly below 1$\,$TeV is perfectly consistent with the preferred di-jet mass of $950\,$GeV of the CMS di-di-jet analysis. We then used the suggested range for $m_X$ from ATLAS to recast the CMS di-di-jet analysis in terms of a resonant search for $Y\!\to\!XX\!\to\!(jj)(jj)$. This significantly reduces the LEE and results in a local (global) significance of $4.0\,\sigma$ $(3.2\,\sigma)$ for a resonance $Y$ with mass $m_Y\approx3.6\TeV$.

We then examined possible combined explanations of the (di-)di-jet excesses and proposed both a model with scalar di-quarks and a model with new heavy colored vector bosons based on an $SU(3)_1\!\times\! SU(3)_2\!\times\! SU(3)_3$ gauge symmetry spontaneously broken to $SU(3)_c$. While the scalar di-quark model has couplings that are at most the order one, the $SU(3)^3$ model requires large non-perturbative couplings, pointing towards an extra-dimensional or composite realization. Interestingly, interpreting this model in a Randall-Sundrum (RS) framework~\cite{Randall:1999ee}, the ratio of the masses of the gauge boson excitations are predicted to be~\cite{Gherghetta:2000qt} 
\begin{equation}
    m_n/m_1=4(n-1/4)/3\,,
\end{equation}
where $m_1$ is the first gluon excitation with a non-vanishing mass. This means if the first resonance ($n=1$) is at $\approx950\,$GeV, the second one ($n=2$) should be at $\approx\!2.2\,$TeV while the third ($n=3$) is at $\approx3.5\,$TeV. While the latter value fits nicely the (di-)di-jet data, this RS framework predicts the existence of another (di-)di-jet resonance with a mass around $2.2\,$TeV. Note that such a resonance, if it has similar couplings to quarks as the $n=1$ and $n=3$ resonances, is not excluded by current di-jet searches due to the PDF scaling w.r.t.~the $950\,$GeV resonance. Furthermore, the CMS di-di-jet data even points towards a slight excess in this region of the di-di-jet invariant mass $m_Y$ (see Appendix).

In light of the intriguing hints for NP in semi-leptonic $B$ decays~\cite{LHCb:2021trn,HFLAV:2019otj}, $g-2$ of the muon~\cite{Bennett:2006fi,Abi:2021gix,Aoyama:2019ryr}, the $W$~mass~\cite{CDF:2022hxs,ParticleDataGroup:2020ssz}, the Cabibbo angle anomaly~\cite{Seng:2022ufm,Manzari:2021kma,Crivellin:2022ctt}, the $96\,$GeV~\cite{CMS-PAS-HIG-17-013}, $151\,$GeV~\cite{Crivellin:2021ubm} and $680\,$GeV~\cite{ATLAS:2017ayi} excesses, the multi-lepton anomalies~\cite{vonBuddenbrock:2016rmr,vonBuddenbrock:2019ajh,vonBuddenbrock:2020ter,Hernandez:2019geu}, the di-Higgs~\cite{ATLAS:2021tyg} excess as well as the hint for non-resonant di-electrons~\cite{CMS:2021ctt,Crivellin:2021rbf},\footnote{See Ref.~\cite{Fischer:2021sqw} for a recent review of anomalies.} the (di-)di-jet excesses constitute one more very interesting sign of physics beyond the SM. While the other signals for NP are in general related electroweak processes within the SM, the (di-)di-jet signals points towards colored new particles. This broadens the range of interactions for which the anomalies suggest NP and has important consequences for collider searches and model building in the collaborative search for the next SM of particle physics.

\medskip
{\it Acknowledgements} --- {We thank Zurab Berezhiani, Monika Blanke, Mukesh Kumar and Xifeng Ruan for useful discussions. B.M.~gratefully acknowledges the support from the South African Department of Science and Innovation through the SA-CERN program and the National Research Foundation for various forms of support. A.C.~is supported by a Professorship Grant (PP00P2\_176884) of the Swiss National Science Foundation.}

\appendix

\section{ATLAS and CMS Plots}

\begin{figure*}
	\centering 
	\includegraphics[width=0.54\textwidth]{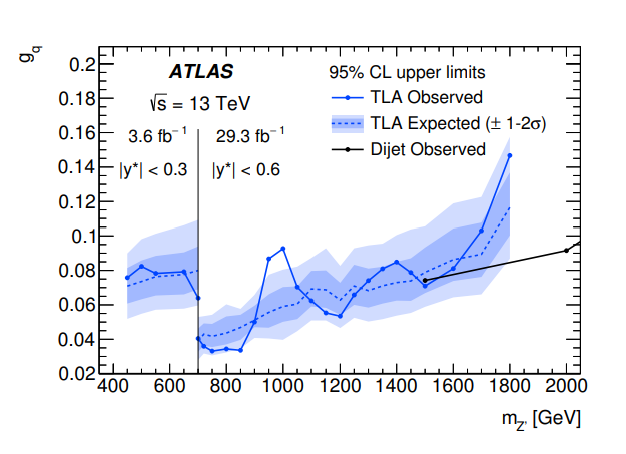}
		\includegraphics[width=0.45\textwidth]{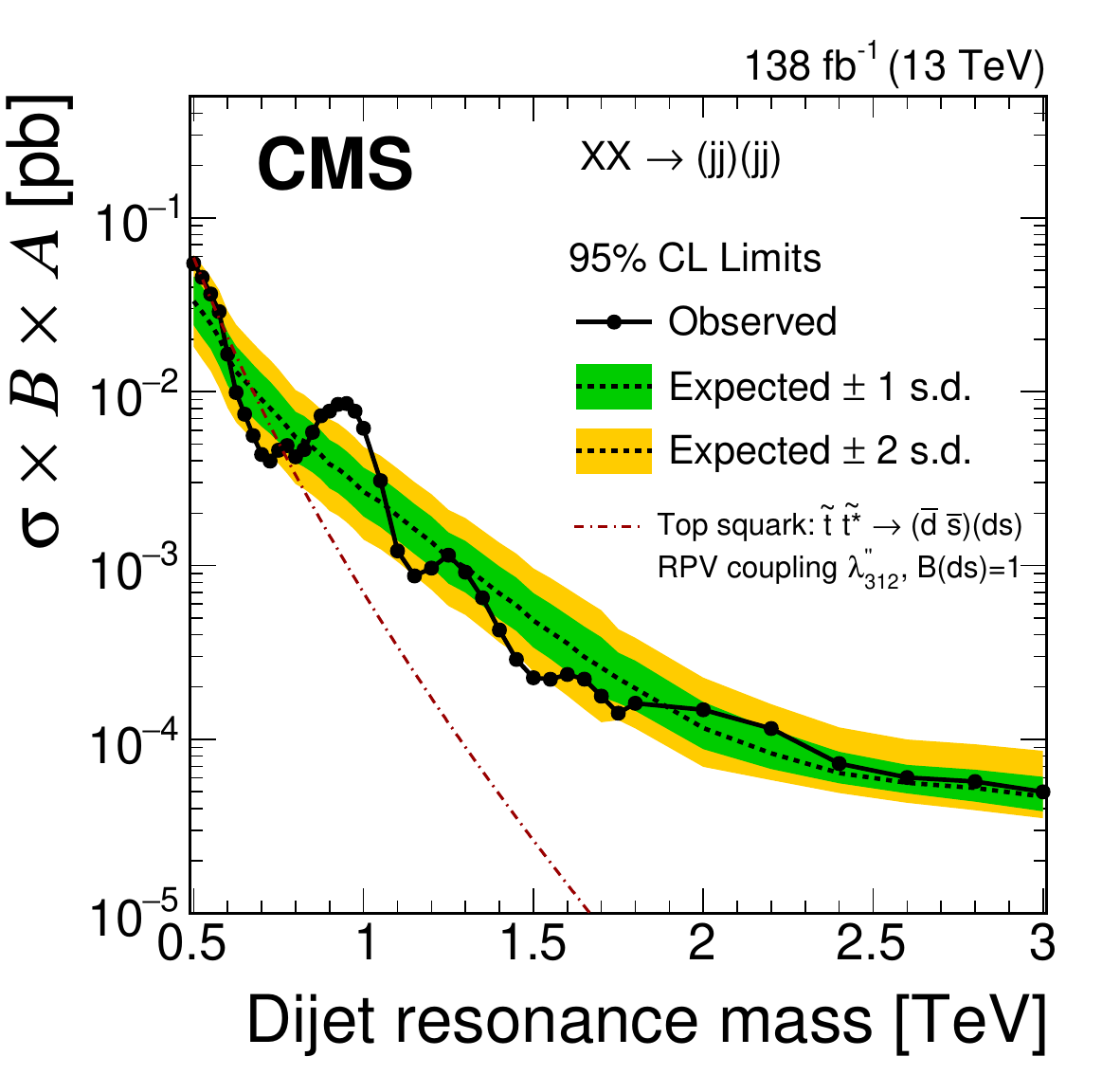}
	\caption{Left: Di-jet search of ATLAS~\cite{ATLAS:2018qto} showing the expected and observed limits on the axial coupling $g_q$ of a $Z^\prime$ boson to quarks. Right: Cross section times branching ratio times acceptance in units of pico barn as a function of the di-jet invariant mass obtained in the CMS di-di-jet analysis~\cite{CMS:2022usq}. }
	\label{fig:dijet}
\end{figure*}

\begin{figure*}
	\centering 
	\includegraphics[width=0.9\textwidth]{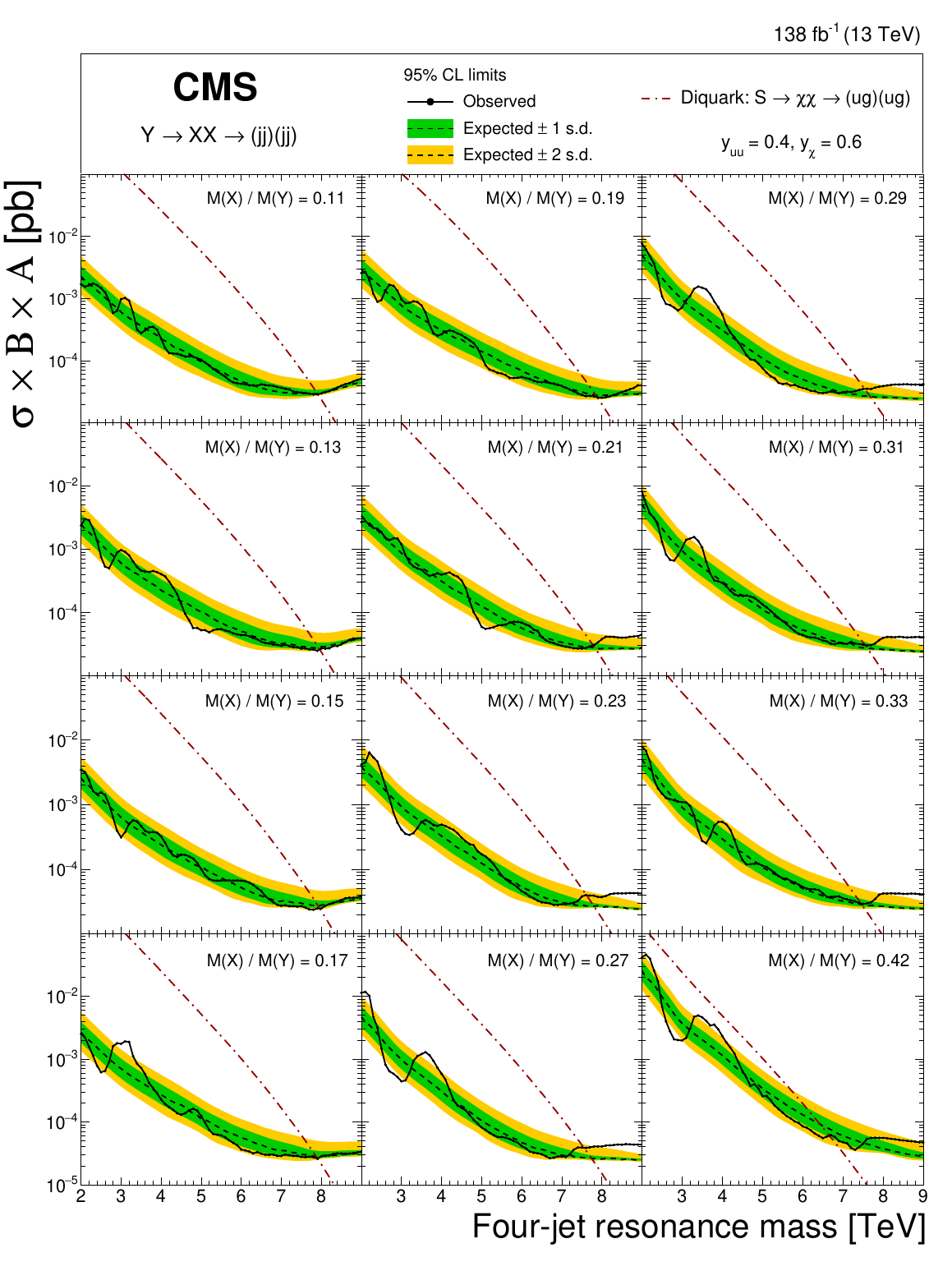}
	\caption{Observed and expected limit on cross section times branching ratio times acceptance in units of pico barn as a function of the di-di-jet invariant mass for different values of $\alpha=m_X/m_Y$~\cite{CMS:2022usq}. }
	\label{fig:didijet}
\end{figure*}

Here we quote the main results of the ATLAS and CMS searches for (di-)di-jet searches for the convenience of the reader. The result of the di-jet resonance search of ATLAS is shown in the left plot of Fig.~\ref{fig:dijet}. The di-jet invariant mass $m_X$ of the CMS di-di-jet analysis is given in the right plot of Fig.~\ref{fig:dijet} while the relevant plots for the di-di-jet mass $m_Y$ are displayed in Fig.~\ref{fig:didijet}.

\bibliography{BIB}

\begin{thebibliography}{71}%
\makeatletter
\providecommand \@ifxundefined [1]{%
 \@ifx{#1\undefined}
}%
\providecommand \@ifnum [1]{%
 \ifnum #1\expandafter \@firstoftwo
 \else \expandafter \@secondoftwo
 \fi
}%
\providecommand \@ifx [1]{%
 \ifx #1\expandafter \@firstoftwo
 \else \expandafter \@secondoftwo
 \fi
}%
\providecommand \natexlab [1]{#1}%
\providecommand \enquote  [1]{``#1''}%
\providecommand \bibnamefont  [1]{#1}%
\providecommand \bibfnamefont [1]{#1}%
\providecommand \citenamefont [1]{#1}%
\providecommand \href@noop [0]{\@secondoftwo}%
\providecommand \href [0]{\begingroup \@sanitize@url \@href}%
\providecommand \@href[1]{\@@startlink{#1}\@@href}%
\providecommand \@@href[1]{\endgroup#1\@@endlink}%
\providecommand \@sanitize@url [0]{\catcode `\\12\catcode `\$12\catcode
  `\&12\catcode `\#12\catcode `\^12\catcode `\_12\catcode `\%12\relax}%
\providecommand \@@startlink[1]{}%
\providecommand \@@endlink[0]{}%
\providecommand \url  [0]{\begingroup\@sanitize@url \@url }%
\providecommand \@url [1]{\endgroup\@href {#1}{\urlprefix }}%
\providecommand \urlprefix  [0]{URL }%
\providecommand \Eprint [0]{\href }%
\providecommand \doibase [0]{https://doi.org/}%
\providecommand \selectlanguage [0]{\@gobble}%
\providecommand \bibinfo  [0]{\@secondoftwo}%
\providecommand \bibfield  [0]{\@secondoftwo}%
\providecommand \translation [1]{[#1]}%
\providecommand \BibitemOpen [0]{}%
\providecommand \bibitemStop [0]{}%
\providecommand \bibitemNoStop [0]{.\EOS\space}%
\providecommand \EOS [0]{\spacefactor3000\relax}%
\providecommand \BibitemShut  [1]{\csname bibitem#1\endcsname}%
\let\auto@bib@innerbib\@empty
%</preamble>
\bibitem [{\citenamefont {Aad}\ \emph {et~al.}(2012)\citenamefont {Aad} \emph
  {et~al.}}]{Aad:2012tfa}%
  \BibitemOpen
  \bibfield  {author} {\bibinfo {author} {\bibfnamefont {G.}~\bibnamefont
  {Aad}} \emph {et~al.} (\bibinfo {collaboration} {ATLAS}),\ }\bibfield
  {title} {\bibinfo {title} {{Observation of a new particle in the search for
  the Standard Model Higgs boson with the ATLAS detector at the LHC}},\ }\href
  {https://doi.org/10.1016/j.physletb.2012.08.020} {\bibfield  {journal}
  {\bibinfo  {journal} {Phys. Lett. B}\ }\textbf {\bibinfo {volume} {716}},\
  \bibinfo {pages} {1} (\bibinfo {year} {2012})},\ \Eprint
  {https://arxiv.org/abs/1207.7214} {arXiv:1207.7214 [hep-ex]} \BibitemShut
  {NoStop}%
\bibitem [{\citenamefont {Chatrchyan}\ \emph {et~al.}(2012)\citenamefont
  {Chatrchyan} \emph {et~al.}}]{Chatrchyan:2012ufa}%
  \BibitemOpen
  \bibfield  {author} {\bibinfo {author} {\bibfnamefont {S.}~\bibnamefont
  {Chatrchyan}} \emph {et~al.} (\bibinfo {collaboration} {CMS}),\ }\bibfield
  {title} {\bibinfo {title} {{Observation of a New Boson at a Mass of 125 GeV
  with the CMS Experiment at the LHC}},\ }\href
  {https://doi.org/10.1016/j.physletb.2012.08.021} {\bibfield  {journal}
  {\bibinfo  {journal} {Phys. Lett. B}\ }\textbf {\bibinfo {volume} {716}},\
  \bibinfo {pages} {30} (\bibinfo {year} {2012})},\ \Eprint
  {https://arxiv.org/abs/1207.7235} {arXiv:1207.7235 [hep-ex]} \BibitemShut
  {NoStop}%
\bibitem [{\citenamefont {Crivellin}\ and\ \citenamefont
  {Matias}(2022)}]{Crivellin:2022qcj}%
  \BibitemOpen
  \bibfield  {author} {\bibinfo {author} {\bibfnamefont {A.}~\bibnamefont
  {Crivellin}}\ and\ \bibinfo {author} {\bibfnamefont {J.}~\bibnamefont
  {Matias}},\ }\bibfield  {title} {\bibinfo {title} {{Beyond the Standard Model
  with Lepton Flavor Universality Violation}},\ }in\ \href@noop {} {\emph
  {\bibinfo {booktitle} {{$1^{\rm st}$ Pan-African Astro-Particle and Collider
  Workshop}}}}\ (\bibinfo {year} {2022})\ \Eprint
  {https://arxiv.org/abs/2204.12175} {arXiv:2204.12175 [hep-ph]} \BibitemShut
  {NoStop}%
\bibitem [{\citenamefont {Crivellin}\ and\ \citenamefont
  {Hoferichter}(2021)}]{Crivellin:2021sff}%
  \BibitemOpen
  \bibfield  {author} {\bibinfo {author} {\bibfnamefont {A.}~\bibnamefont
  {Crivellin}}\ and\ \bibinfo {author} {\bibfnamefont {M.}~\bibnamefont
  {Hoferichter}},\ }\bibfield  {title} {\bibinfo {title} {{Hints of lepton
  flavor universality violations}},\ }\href
  {https://doi.org/10.1126/science.abk2450} {\bibfield  {journal} {\bibinfo
  {journal} {Science}\ }\textbf {\bibinfo {volume} {374}},\ \bibinfo {pages}
  {1051} (\bibinfo {year} {2021})},\ \Eprint {https://arxiv.org/abs/2111.12739}
  {arXiv:2111.12739 [hep-ph]} \BibitemShut {NoStop}%
\bibitem [{\citenamefont {Fischer}\ \emph {et~al.}(2022)\citenamefont {Fischer}
  \emph {et~al.}}]{Fischer:2021sqw}%
  \BibitemOpen
  \bibfield  {author} {\bibinfo {author} {\bibfnamefont {O.}~\bibnamefont
  {Fischer}} \emph {et~al.},\ }\bibfield  {title} {\bibinfo {title} {{Unveiling
  hidden physics at the LHC}},\ }\href
  {https://doi.org/10.1140/epjc/s10052-022-10541-4} {\bibfield  {journal}
  {\bibinfo  {journal} {Eur. Phys. J. C}\ }\textbf {\bibinfo {volume} {82}},\
  \bibinfo {pages} {665} (\bibinfo {year} {2022})},\ \Eprint
  {https://arxiv.org/abs/2109.06065} {arXiv:2109.06065 [hep-ph]} \BibitemShut
  {NoStop}%
\bibitem [{\citenamefont {Aaboud}\ \emph {et~al.}(2018)\citenamefont {Aaboud}
  \emph {et~al.}}]{ATLAS:2018qto}%
  \BibitemOpen
  \bibfield  {author} {\bibinfo {author} {\bibfnamefont {M.}~\bibnamefont
  {Aaboud}} \emph {et~al.} (\bibinfo {collaboration} {ATLAS}),\ }\bibfield
  {title} {\bibinfo {title} {{Search for low-mass dijet resonances using
  trigger-level jets with the ATLAS detector in $pp$ collisions at
  $\sqrt{s}=13$ TeV}},\ }\href {https://doi.org/10.1103/PhysRevLett.121.081801}
  {\bibfield  {journal} {\bibinfo  {journal} {Phys. Rev. Lett.}\ }\textbf
  {\bibinfo {volume} {121}},\ \bibinfo {pages} {081801} (\bibinfo {year}
  {2018})},\ \Eprint {https://arxiv.org/abs/1804.03496} {arXiv:1804.03496
  [hep-ex]} \BibitemShut {NoStop}%
\bibitem [{\citenamefont {Sirunyan}\ \emph {et~al.}(2018)\citenamefont
  {Sirunyan} \emph {et~al.}}]{CMS:2018mgb}%
  \BibitemOpen
  \bibfield  {author} {\bibinfo {author} {\bibfnamefont {A.~M.}\ \bibnamefont
  {Sirunyan}} \emph {et~al.} (\bibinfo {collaboration} {CMS}),\ }\bibfield
  {title} {\bibinfo {title} {{Search for narrow and broad dijet resonances in
  proton-proton collisions at $ \sqrt{s}=13 $ TeV and constraints on dark
  matter mediators and other new particles}},\ }\href
  {https://doi.org/10.1007/JHEP08(2018)130} {\bibfield  {journal} {\bibinfo
  {journal} {JHEP}\ }\textbf {\bibinfo {volume} {08}},\ \bibinfo {pages}
  {130}},\ \Eprint {https://arxiv.org/abs/1806.00843} {arXiv:1806.00843
  [hep-ex]} \BibitemShut {NoStop}%
\bibitem [{CMS(2022)}]{CMS:2022usq}%
  \BibitemOpen
  \href@noop {} {\bibinfo {title} {{Search for resonant and nonresonant
  production of pairs of dijet resonances in proton-proton collisions at
  $\sqrt{s}$ = 13 TeV}}} (\bibinfo {year} {2022}),\ \Eprint
  {https://arxiv.org/abs/2206.09997} {arXiv:2206.09997 [hep-ex]} \BibitemShut
  {NoStop}%
\bibitem [{\citenamefont {Han}\ \emph {et~al.}(2010)\citenamefont {Han},
  \citenamefont {Lewis},\ and\ \citenamefont {McElmurry}}]{Han:2009ya}%
  \BibitemOpen
  \bibfield  {author} {\bibinfo {author} {\bibfnamefont {T.}~\bibnamefont
  {Han}}, \bibinfo {author} {\bibfnamefont {I.}~\bibnamefont {Lewis}},\ and\
  \bibinfo {author} {\bibfnamefont {T.}~\bibnamefont {McElmurry}},\ }\bibfield
  {title} {\bibinfo {title} {{QCD Corrections to Scalar Diquark Production at
  Hadron Colliders}},\ }\href {https://doi.org/10.1007/JHEP01(2010)123}
  {\bibfield  {journal} {\bibinfo  {journal} {JHEP}\ }\textbf {\bibinfo
  {volume} {01}},\ \bibinfo {pages} {123}},\ \Eprint
  {https://arxiv.org/abs/0909.2666} {arXiv:0909.2666 [hep-ph]} \BibitemShut
  {NoStop}%
\bibitem [{\citenamefont {Gogoladze}\ \emph {et~al.}(2010)\citenamefont
  {Gogoladze}, \citenamefont {Mimura}, \citenamefont {Okada},\ and\
  \citenamefont {Shafi}}]{Gogoladze:2010xd}%
  \BibitemOpen
  \bibfield  {author} {\bibinfo {author} {\bibfnamefont {I.}~\bibnamefont
  {Gogoladze}}, \bibinfo {author} {\bibfnamefont {Y.}~\bibnamefont {Mimura}},
  \bibinfo {author} {\bibfnamefont {N.}~\bibnamefont {Okada}},\ and\ \bibinfo
  {author} {\bibfnamefont {Q.}~\bibnamefont {Shafi}},\ }\bibfield  {title}
  {\bibinfo {title} {{Color Triplet Diquarks at the LHC}},\ }\href
  {https://doi.org/10.1016/j.physletb.2010.02.068} {\bibfield  {journal}
  {\bibinfo  {journal} {Phys. Lett. B}\ }\textbf {\bibinfo {volume} {686}},\
  \bibinfo {pages} {233} (\bibinfo {year} {2010})},\ \Eprint
  {https://arxiv.org/abs/1001.5260} {arXiv:1001.5260 [hep-ph]} \BibitemShut
  {NoStop}%
\bibitem [{\citenamefont {Giudice}\ \emph {et~al.}(2011)\citenamefont
  {Giudice}, \citenamefont {Gripaios},\ and\ \citenamefont
  {Sundrum}}]{Giudice:2011ak}%
  \BibitemOpen
  \bibfield  {author} {\bibinfo {author} {\bibfnamefont {G.~F.}\ \bibnamefont
  {Giudice}}, \bibinfo {author} {\bibfnamefont {B.}~\bibnamefont {Gripaios}},\
  and\ \bibinfo {author} {\bibfnamefont {R.}~\bibnamefont {Sundrum}},\
  }\bibfield  {title} {\bibinfo {title} {{Flavourful Production at Hadron
  Colliders}},\ }\href {https://doi.org/10.1007/JHEP08(2011)055} {\bibfield
  {journal} {\bibinfo  {journal} {JHEP}\ }\textbf {\bibinfo {volume} {08}},\
  \bibinfo {pages} {055}},\ \Eprint {https://arxiv.org/abs/1105.3161}
  {arXiv:1105.3161 [hep-ph]} \BibitemShut {NoStop}%
\bibitem [{\citenamefont {Sirunyan}\ \emph {et~al.}(2020)\citenamefont
  {Sirunyan} \emph {et~al.}}]{CMS:2019gwf}%
  \BibitemOpen
  \bibfield  {author} {\bibinfo {author} {\bibfnamefont {A.~M.}\ \bibnamefont
  {Sirunyan}} \emph {et~al.} (\bibinfo {collaboration} {CMS}),\ }\bibfield
  {title} {\bibinfo {title} {{Search for high mass dijet resonances with a new
  background prediction method in proton-proton collisions at $\sqrt{s} =$ 13
  TeV}},\ }\href {https://doi.org/10.1007/JHEP05(2020)033} {\bibfield
  {journal} {\bibinfo  {journal} {JHEP}\ }\textbf {\bibinfo {volume} {05}},\
  \bibinfo {pages} {033}},\ \Eprint {https://arxiv.org/abs/1911.03947}
  {arXiv:1911.03947 [hep-ex]} \BibitemShut {NoStop}%
\bibitem [{\citenamefont {Aad}\ \emph {et~al.}(2020{\natexlab{a}})\citenamefont
  {Aad} \emph {et~al.}}]{ATLAS:2019fgd}%
  \BibitemOpen
  \bibfield  {author} {\bibinfo {author} {\bibfnamefont {G.}~\bibnamefont
  {Aad}} \emph {et~al.} (\bibinfo {collaboration} {ATLAS}),\ }\bibfield
  {title} {\bibinfo {title} {{Search for new resonances in mass distributions
  of jet pairs using 139 fb$^{-1}$ of $pp$ collisions at $\sqrt{s}=13$ TeV with
  the ATLAS detector}},\ }\href {https://doi.org/10.1007/JHEP03(2020)145}
  {\bibfield  {journal} {\bibinfo  {journal} {JHEP}\ }\textbf {\bibinfo
  {volume} {03}},\ \bibinfo {pages} {145}},\ \Eprint
  {https://arxiv.org/abs/1910.08447} {arXiv:1910.08447 [hep-ex]} \BibitemShut
  {NoStop}%
\bibitem [{\citenamefont {Alwall}\ \emph {et~al.}(2014)\citenamefont {Alwall},
  \citenamefont {Frederix}, \citenamefont {Frixione}, \citenamefont {Hirschi},
  \citenamefont {Maltoni}, \citenamefont {Mattelaer}, \citenamefont {Shao},
  \citenamefont {Stelzer}, \citenamefont {Torrielli},\ and\ \citenamefont
  {Zaro}}]{Alwall:2014hca}%
  \BibitemOpen
  \bibfield  {author} {\bibinfo {author} {\bibfnamefont {J.}~\bibnamefont
  {Alwall}}, \bibinfo {author} {\bibfnamefont {R.}~\bibnamefont {Frederix}},
  \bibinfo {author} {\bibfnamefont {S.}~\bibnamefont {Frixione}}, \bibinfo
  {author} {\bibfnamefont {V.}~\bibnamefont {Hirschi}}, \bibinfo {author}
  {\bibfnamefont {F.}~\bibnamefont {Maltoni}}, \bibinfo {author} {\bibfnamefont
  {O.}~\bibnamefont {Mattelaer}}, \bibinfo {author} {\bibfnamefont {H.~S.}\
  \bibnamefont {Shao}}, \bibinfo {author} {\bibfnamefont {T.}~\bibnamefont
  {Stelzer}}, \bibinfo {author} {\bibfnamefont {P.}~\bibnamefont {Torrielli}},\
  and\ \bibinfo {author} {\bibfnamefont {M.}~\bibnamefont {Zaro}},\ }\bibfield
  {title} {\bibinfo {title} {{The automated computation of tree-level and
  next-to-leading order differential cross sections, and their matching to
  parton shower simulations}},\ }\href
  {https://doi.org/10.1007/JHEP07(2014)079} {\bibfield  {journal} {\bibinfo
  {journal} {JHEP}\ }\textbf {\bibinfo {volume} {07}},\ \bibinfo {pages}
  {079}},\ \Eprint {https://arxiv.org/abs/1405.0301} {arXiv:1405.0301 [hep-ph]}
  \BibitemShut {NoStop}%
\bibitem [{\citenamefont {Sj\"ostrand}\ \emph {et~al.}(2015)\citenamefont
  {Sj\"ostrand}, \citenamefont {Ask}, \citenamefont {Christiansen},
  \citenamefont {Corke}, \citenamefont {Desai}, \citenamefont {Ilten},
  \citenamefont {Mrenna}, \citenamefont {Prestel}, \citenamefont {Rasmussen},\
  and\ \citenamefont {Skands}}]{Sjostrand:2014zea}%
  \BibitemOpen
  \bibfield  {author} {\bibinfo {author} {\bibfnamefont {T.}~\bibnamefont
  {Sj\"ostrand}}, \bibinfo {author} {\bibfnamefont {S.}~\bibnamefont {Ask}},
  \bibinfo {author} {\bibfnamefont {J.~R.}\ \bibnamefont {Christiansen}},
  \bibinfo {author} {\bibfnamefont {R.}~\bibnamefont {Corke}}, \bibinfo
  {author} {\bibfnamefont {N.}~\bibnamefont {Desai}}, \bibinfo {author}
  {\bibfnamefont {P.}~\bibnamefont {Ilten}}, \bibinfo {author} {\bibfnamefont
  {S.}~\bibnamefont {Mrenna}}, \bibinfo {author} {\bibfnamefont
  {S.}~\bibnamefont {Prestel}}, \bibinfo {author} {\bibfnamefont {C.~O.}\
  \bibnamefont {Rasmussen}},\ and\ \bibinfo {author} {\bibfnamefont {P.~Z.}\
  \bibnamefont {Skands}},\ }\bibfield  {title} {\bibinfo {title} {{An
  introduction to PYTHIA 8.2}},\ }\href
  {https://doi.org/10.1016/j.cpc.2015.01.024} {\bibfield  {journal} {\bibinfo
  {journal} {Comput. Phys. Commun.}\ }\textbf {\bibinfo {volume} {191}},\
  \bibinfo {pages} {159} (\bibinfo {year} {2015})},\ \Eprint
  {https://arxiv.org/abs/1410.3012} {arXiv:1410.3012 [hep-ph]} \BibitemShut
  {NoStop}%
\bibitem [{\citenamefont {Ball}\ \emph
  {et~al.}(2013{\natexlab{a}})\citenamefont {Ball} \emph
  {et~al.}}]{Ball:2012cx}%
  \BibitemOpen
  \bibfield  {author} {\bibinfo {author} {\bibfnamefont {R.~D.}\ \bibnamefont
  {Ball}} \emph {et~al.},\ }\bibfield  {title} {\bibinfo {title} {{Parton
  distributions with LHC data}},\ }\href
  {https://doi.org/10.1016/j.nuclphysb.2012.10.003} {\bibfield  {journal}
  {\bibinfo  {journal} {Nucl. Phys. B}\ }\textbf {\bibinfo {volume} {867}},\
  \bibinfo {pages} {244} (\bibinfo {year} {2013}{\natexlab{a}})},\ \Eprint
  {https://arxiv.org/abs/1207.1303} {arXiv:1207.1303 [hep-ph]} \BibitemShut
  {NoStop}%
\bibitem [{\citenamefont {de~Favereau}\ \emph {et~al.}(2014)\citenamefont
  {de~Favereau}, \citenamefont {Delaere}, \citenamefont {Demin}, \citenamefont
  {Giammanco}, \citenamefont {Lema\^\i{}tre}, \citenamefont {Mertens},\ and\
  \citenamefont {Selvaggi}}]{deFavereau:2013fsa}%
  \BibitemOpen
  \bibfield  {author} {\bibinfo {author} {\bibfnamefont {J.}~\bibnamefont
  {de~Favereau}}, \bibinfo {author} {\bibfnamefont {C.}~\bibnamefont
  {Delaere}}, \bibinfo {author} {\bibfnamefont {P.}~\bibnamefont {Demin}},
  \bibinfo {author} {\bibfnamefont {A.}~\bibnamefont {Giammanco}}, \bibinfo
  {author} {\bibfnamefont {V.}~\bibnamefont {Lema\^\i{}tre}}, \bibinfo {author}
  {\bibfnamefont {A.}~\bibnamefont {Mertens}},\ and\ \bibinfo {author}
  {\bibfnamefont {M.}~\bibnamefont {Selvaggi}} (\bibinfo {collaboration}
  {DELPHES 3}),\ }\bibfield  {title} {\bibinfo {title} {{DELPHES 3, A modular
  framework for fast simulation of a generic collider experiment}},\ }\href
  {https://doi.org/10.1007/JHEP02(2014)057} {\bibfield  {journal} {\bibinfo
  {journal} {JHEP}\ }\textbf {\bibinfo {volume} {02}},\ \bibinfo {pages}
  {057}},\ \Eprint {https://arxiv.org/abs/1307.6346} {arXiv:1307.6346 [hep-ex]}
  \BibitemShut {NoStop}%
\bibitem [{\citenamefont {Cacciari}\ \emph {et~al.}(2008)\citenamefont
  {Cacciari}, \citenamefont {Salam},\ and\ \citenamefont
  {Soyez}}]{Cacciari:2008gp}%
  \BibitemOpen
  \bibfield  {author} {\bibinfo {author} {\bibfnamefont {M.}~\bibnamefont
  {Cacciari}}, \bibinfo {author} {\bibfnamefont {G.~P.}\ \bibnamefont
  {Salam}},\ and\ \bibinfo {author} {\bibfnamefont {G.}~\bibnamefont {Soyez}},\
  }\bibfield  {title} {\bibinfo {title} {{The anti-$k_t$ jet clustering
  algorithm}},\ }\href {https://doi.org/10.1088/1126-6708/2008/04/063}
  {\bibfield  {journal} {\bibinfo  {journal} {JHEP}\ }\textbf {\bibinfo
  {volume} {04}},\ \bibinfo {pages} {063}},\ \Eprint
  {https://arxiv.org/abs/0802.1189} {arXiv:0802.1189 [hep-ph]} \BibitemShut
  {NoStop}%
\bibitem [{\citenamefont {Cacciari}\ \emph {et~al.}(2012)\citenamefont
  {Cacciari}, \citenamefont {Salam},\ and\ \citenamefont
  {Soyez}}]{Cacciari:2011ma}%
  \BibitemOpen
  \bibfield  {author} {\bibinfo {author} {\bibfnamefont {M.}~\bibnamefont
  {Cacciari}}, \bibinfo {author} {\bibfnamefont {G.~P.}\ \bibnamefont
  {Salam}},\ and\ \bibinfo {author} {\bibfnamefont {G.}~\bibnamefont {Soyez}},\
  }\bibfield  {title} {\bibinfo {title} {{FastJet User Manual}},\ }\href
  {https://doi.org/10.1140/epjc/s10052-012-1896-2} {\bibfield  {journal}
  {\bibinfo  {journal} {Eur. Phys. J. C}\ }\textbf {\bibinfo {volume} {72}},\
  \bibinfo {pages} {1896} (\bibinfo {year} {2012})},\ \Eprint
  {https://arxiv.org/abs/1111.6097} {arXiv:1111.6097 [hep-ph]} \BibitemShut
  {NoStop}%
\bibitem [{\citenamefont {Lyons}(2020)}]{Lyons2020}%
  \BibitemOpen
  \bibfield  {author} {\bibinfo {author} {\bibfnamefont {L.}~\bibnamefont
  {Lyons}},\ }\bibinfo {title} {Statistical issues in particle physics},\ in\
  \href {https://doi.org/10.1007/978-3-030-35318-6_15} {\emph {\bibinfo
  {booktitle} {Particle Physics Reference Library}}},\ Vol.\ \bibinfo {volume}
  {2: Detectors for Particles and Radiation},\ \bibinfo {editor} {edited by\
  \bibinfo {editor} {\bibfnamefont {C.~W.}\ \bibnamefont {Fabjan}}\ and\
  \bibinfo {editor} {\bibfnamefont {H.}~\bibnamefont {Schopper}}}\ (\bibinfo
  {publisher} {Springer International Publishing},\ \bibinfo {address} {Cham},\
  \bibinfo {year} {2020})\ pp.\ \bibinfo {pages} {645--692}\BibitemShut
  {NoStop}%
\bibitem [{\citenamefont {Aad}\ \emph {et~al.}(2020{\natexlab{b}})\citenamefont
  {Aad} \emph {et~al.}}]{ATLAS:2020yat}%
  \BibitemOpen
  \bibfield  {author} {\bibinfo {author} {\bibfnamefont {G.}~\bibnamefont
  {Aad}} \emph {et~al.} (\bibinfo {collaboration} {ATLAS}),\ }\bibfield
  {title} {\bibinfo {title} {{Search for new non-resonant phenomena in
  high-mass dilepton final states with the ATLAS detector}},\ }\href
  {https://doi.org/10.1007/JHEP11(2020)005} {\bibfield  {journal} {\bibinfo
  {journal} {JHEP}\ }\textbf {\bibinfo {volume} {11}},\ \bibinfo {pages}
  {005}},\ \bibinfo {note} {[Erratum: JHEP 04, 142 (2021)]},\ \Eprint
  {https://arxiv.org/abs/2006.12946} {arXiv:2006.12946 [hep-ex]} \BibitemShut
  {NoStop}%
\bibitem [{\citenamefont {Sirunyan}\ \emph {et~al.}(2021)\citenamefont
  {Sirunyan} \emph {et~al.}}]{CMS:2021ctt}%
  \BibitemOpen
  \bibfield  {author} {\bibinfo {author} {\bibfnamefont {A.~M.}\ \bibnamefont
  {Sirunyan}} \emph {et~al.} (\bibinfo {collaboration} {CMS}),\ }\bibfield
  {title} {\bibinfo {title} {{Search for resonant and nonresonant new phenomena
  in high-mass dilepton final states at $ \sqrt{s} $ = 13 TeV}},\ }\href
  {https://doi.org/10.1007/JHEP07(2021)208} {\bibfield  {journal} {\bibinfo
  {journal} {JHEP}\ }\textbf {\bibinfo {volume} {07}},\ \bibinfo {pages}
  {208}},\ \Eprint {https://arxiv.org/abs/2103.02708} {arXiv:2103.02708
  [hep-ex]} \BibitemShut {NoStop}%
\bibitem [{\citenamefont {Schael}\ \emph {et~al.}(2006)\citenamefont {Schael}
  \emph {et~al.}}]{ALEPH:2005ab}%
  \BibitemOpen
  \bibfield  {author} {\bibinfo {author} {\bibfnamefont {S.}~\bibnamefont
  {Schael}} \emph {et~al.} (\bibinfo {collaboration} {ALEPH, DELPHI, L3, OPAL,
  SLD, LEP Electroweak Working Group, SLD Electroweak Group, SLD Heavy Flavour
  Group}),\ }\bibfield  {title} {\bibinfo {title} {{Precision electroweak
  measurements on the $Z$ resonance}},\ }\href
  {https://doi.org/10.1016/j.physrep.2005.12.006} {\bibfield  {journal}
  {\bibinfo  {journal} {Phys. Rept.}\ }\textbf {\bibinfo {volume} {427}},\
  \bibinfo {pages} {257} (\bibinfo {year} {2006})},\ \Eprint
  {https://arxiv.org/abs/hep-ex/0509008} {arXiv:hep-ex/0509008} \BibitemShut
  {NoStop}%
\bibitem [{\citenamefont {Schael}\ \emph {et~al.}(2013)\citenamefont {Schael}
  \emph {et~al.}}]{ALEPH:2013dgf}%
  \BibitemOpen
  \bibfield  {author} {\bibinfo {author} {\bibfnamefont {S.}~\bibnamefont
  {Schael}} \emph {et~al.} (\bibinfo {collaboration} {ALEPH, DELPHI, L3, OPAL,
  LEP Electroweak}),\ }\bibfield  {title} {\bibinfo {title} {{Electroweak
  Measurements in Electron-Positron Collisions at W-Boson-Pair Energies at
  LEP}},\ }\href {https://doi.org/10.1016/j.physrep.2013.07.004} {\bibfield
  {journal} {\bibinfo  {journal} {Phys. Rept.}\ }\textbf {\bibinfo {volume}
  {532}},\ \bibinfo {pages} {119} (\bibinfo {year} {2013})},\ \Eprint
  {https://arxiv.org/abs/1302.3415} {arXiv:1302.3415 [hep-ex]} \BibitemShut
  {NoStop}%
\bibitem [{\citenamefont {Hill}(1991)}]{Hill:1991at}%
  \BibitemOpen
  \bibfield  {author} {\bibinfo {author} {\bibfnamefont {C.~T.}\ \bibnamefont
  {Hill}},\ }\bibfield  {title} {\bibinfo {title} {{Topcolor: Top quark
  condensation in a gauge extension of the standard model}},\ }\href
  {https://doi.org/10.1016/0370-2693(91)91061-Y} {\bibfield  {journal}
  {\bibinfo  {journal} {Phys. Lett. B}\ }\textbf {\bibinfo {volume} {266}},\
  \bibinfo {pages} {419} (\bibinfo {year} {1991})}\BibitemShut {NoStop}%
\bibitem [{\citenamefont {Frampton}\ and\ \citenamefont
  {Glashow}(1987)}]{Frampton:1987dn}%
  \BibitemOpen
  \bibfield  {author} {\bibinfo {author} {\bibfnamefont {P.~H.}\ \bibnamefont
  {Frampton}}\ and\ \bibinfo {author} {\bibfnamefont {S.~L.}\ \bibnamefont
  {Glashow}},\ }\bibfield  {title} {\bibinfo {title} {{Chiral Color: An
  Alternative to the Standard Model}},\ }\href
  {https://doi.org/10.1016/0370-2693(87)90859-8} {\bibfield  {journal}
  {\bibinfo  {journal} {Phys. Lett. B}\ }\textbf {\bibinfo {volume} {190}},\
  \bibinfo {pages} {157} (\bibinfo {year} {1987})}\BibitemShut {NoStop}%
\bibitem [{\citenamefont {Chivukula}\ \emph {et~al.}(1996)\citenamefont
  {Chivukula}, \citenamefont {Cohen},\ and\ \citenamefont
  {Simmons}}]{Chivukula:1996yr}%
  \BibitemOpen
  \bibfield  {author} {\bibinfo {author} {\bibfnamefont {R.~S.}\ \bibnamefont
  {Chivukula}}, \bibinfo {author} {\bibfnamefont {A.~G.}\ \bibnamefont
  {Cohen}},\ and\ \bibinfo {author} {\bibfnamefont {E.~H.}\ \bibnamefont
  {Simmons}},\ }\bibfield  {title} {\bibinfo {title} {{New strong interactions
  at the Tevatron?}},\ }\href {https://doi.org/10.1016/0370-2693(96)00464-9}
  {\bibfield  {journal} {\bibinfo  {journal} {Phys. Lett. B}\ }\textbf
  {\bibinfo {volume} {380}},\ \bibinfo {pages} {92} (\bibinfo {year} {1996})},\
  \Eprint {https://arxiv.org/abs/hep-ph/9603311} {arXiv:hep-ph/9603311}
  \BibitemShut {NoStop}%
\bibitem [{\citenamefont {Martynov}\ and\ \citenamefont
  {Smirnov}(2009)}]{Martynov:2009en}%
  \BibitemOpen
  \bibfield  {author} {\bibinfo {author} {\bibfnamefont {M.~V.}\ \bibnamefont
  {Martynov}}\ and\ \bibinfo {author} {\bibfnamefont {A.~D.}\ \bibnamefont
  {Smirnov}},\ }\bibfield  {title} {\bibinfo {title} {{Chiral color symmetry
  and possible G-prime-boson effects at the Tevatron and LHC}},\ }\href
  {https://doi.org/10.1142/S0217732309031296} {\bibfield  {journal} {\bibinfo
  {journal} {Mod. Phys. Lett. A}\ }\textbf {\bibinfo {volume} {24}},\ \bibinfo
  {pages} {1897} (\bibinfo {year} {2009})},\ \Eprint
  {https://arxiv.org/abs/0906.4525} {arXiv:0906.4525 [hep-ph]} \BibitemShut
  {NoStop}%
\bibitem [{\citenamefont {Frampton}\ \emph {et~al.}(2010)\citenamefont
  {Frampton}, \citenamefont {Shu},\ and\ \citenamefont
  {Wang}}]{Frampton:2009rk}%
  \BibitemOpen
  \bibfield  {author} {\bibinfo {author} {\bibfnamefont {P.~H.}\ \bibnamefont
  {Frampton}}, \bibinfo {author} {\bibfnamefont {J.}~\bibnamefont {Shu}},\ and\
  \bibinfo {author} {\bibfnamefont {K.}~\bibnamefont {Wang}},\ }\bibfield
  {title} {\bibinfo {title} {{Axigluon as Possible Explanation for p anti-p
  ---\ensuremath{>} t anti-t Forward-Backward Asymmetry}},\ }\href
  {https://doi.org/10.1016/j.physletb.2009.12.043} {\bibfield  {journal}
  {\bibinfo  {journal} {Phys. Lett. B}\ }\textbf {\bibinfo {volume} {683}},\
  \bibinfo {pages} {294} (\bibinfo {year} {2010})},\ \Eprint
  {https://arxiv.org/abs/0911.2955} {arXiv:0911.2955 [hep-ph]} \BibitemShut
  {NoStop}%
\bibitem [{\citenamefont {Chivukula}\ \emph {et~al.}(2013)\citenamefont
  {Chivukula}, \citenamefont {Simmons},\ and\ \citenamefont
  {Vignaroli}}]{Chivukula:2013kw}%
  \BibitemOpen
  \bibfield  {author} {\bibinfo {author} {\bibfnamefont {R.~S.}\ \bibnamefont
  {Chivukula}}, \bibinfo {author} {\bibfnamefont {E.~H.}\ \bibnamefont
  {Simmons}},\ and\ \bibinfo {author} {\bibfnamefont {N.}~\bibnamefont
  {Vignaroli}},\ }\bibfield  {title} {\bibinfo {title} {{A Flavorful
  Top-Coloron Model}},\ }\href {https://doi.org/10.1103/PhysRevD.87.075002}
  {\bibfield  {journal} {\bibinfo  {journal} {Phys. Rev. D}\ }\textbf {\bibinfo
  {volume} {87}},\ \bibinfo {pages} {075002} (\bibinfo {year} {2013})},\
  \Eprint {https://arxiv.org/abs/1302.1069} {arXiv:1302.1069 [hep-ph]}
  \BibitemShut {NoStop}%
\bibitem [{\citenamefont {Han}\ \emph {et~al.}(1999)\citenamefont {Han},
  \citenamefont {Lykken},\ and\ \citenamefont {Zhang}}]{Han:1998sg}%
  \BibitemOpen
  \bibfield  {author} {\bibinfo {author} {\bibfnamefont {T.}~\bibnamefont
  {Han}}, \bibinfo {author} {\bibfnamefont {J.~D.}\ \bibnamefont {Lykken}},\
  and\ \bibinfo {author} {\bibfnamefont {R.-J.}\ \bibnamefont {Zhang}},\
  }\bibfield  {title} {\bibinfo {title} {{On Kaluza-Klein states from large
  extra dimensions}},\ }\href {https://doi.org/10.1103/PhysRevD.59.105006}
  {\bibfield  {journal} {\bibinfo  {journal} {Phys. Rev. D}\ }\textbf {\bibinfo
  {volume} {59}},\ \bibinfo {pages} {105006} (\bibinfo {year} {1999})},\
  \Eprint {https://arxiv.org/abs/hep-ph/9811350} {arXiv:hep-ph/9811350}
  \BibitemShut {NoStop}%
\bibitem [{\citenamefont {Dicus}\ \emph {et~al.}(2002)\citenamefont {Dicus},
  \citenamefont {McMullen},\ and\ \citenamefont {Nandi}}]{Dicus:2000hm}%
  \BibitemOpen
  \bibfield  {author} {\bibinfo {author} {\bibfnamefont {D.~A.}\ \bibnamefont
  {Dicus}}, \bibinfo {author} {\bibfnamefont {C.~D.}\ \bibnamefont
  {McMullen}},\ and\ \bibinfo {author} {\bibfnamefont {S.}~\bibnamefont
  {Nandi}},\ }\bibfield  {title} {\bibinfo {title} {{Collider implications of
  Kaluza-Klein excitations of the gluons}},\ }\href
  {https://doi.org/10.1103/PhysRevD.65.076007} {\bibfield  {journal} {\bibinfo
  {journal} {Phys. Rev. D}\ }\textbf {\bibinfo {volume} {65}},\ \bibinfo
  {pages} {076007} (\bibinfo {year} {2002})},\ \Eprint
  {https://arxiv.org/abs/hep-ph/0012259} {arXiv:hep-ph/0012259} \BibitemShut
  {NoStop}%
\bibitem [{\citenamefont {Davoudiasl}\ \emph {et~al.}(2001)\citenamefont
  {Davoudiasl}, \citenamefont {Hewett},\ and\ \citenamefont
  {Rizzo}}]{Davoudiasl:2000wi}%
  \BibitemOpen
  \bibfield  {author} {\bibinfo {author} {\bibfnamefont {H.}~\bibnamefont
  {Davoudiasl}}, \bibinfo {author} {\bibfnamefont {J.~L.}\ \bibnamefont
  {Hewett}},\ and\ \bibinfo {author} {\bibfnamefont {T.~G.}\ \bibnamefont
  {Rizzo}},\ }\bibfield  {title} {\bibinfo {title} {{Experimental probes of
  localized gravity: On and off the wall}},\ }\href
  {https://doi.org/10.1103/PhysRevD.63.075004} {\bibfield  {journal} {\bibinfo
  {journal} {Phys. Rev. D}\ }\textbf {\bibinfo {volume} {63}},\ \bibinfo
  {pages} {075004} (\bibinfo {year} {2001})},\ \Eprint
  {https://arxiv.org/abs/hep-ph/0006041} {arXiv:hep-ph/0006041} \BibitemShut
  {NoStop}%
\bibitem [{\citenamefont {Lillie}\ \emph {et~al.}(2007)\citenamefont {Lillie},
  \citenamefont {Randall},\ and\ \citenamefont {Wang}}]{Lillie:2007yh}%
  \BibitemOpen
  \bibfield  {author} {\bibinfo {author} {\bibfnamefont {B.}~\bibnamefont
  {Lillie}}, \bibinfo {author} {\bibfnamefont {L.}~\bibnamefont {Randall}},\
  and\ \bibinfo {author} {\bibfnamefont {L.-T.}\ \bibnamefont {Wang}},\
  }\bibfield  {title} {\bibinfo {title} {{The Bulk RS KK-gluon at the LHC}},\
  }\href {https://doi.org/10.1088/1126-6708/2007/09/074} {\bibfield  {journal}
  {\bibinfo  {journal} {JHEP}\ }\textbf {\bibinfo {volume} {09}},\ \bibinfo
  {pages} {074}},\ \Eprint {https://arxiv.org/abs/hep-ph/0701166}
  {arXiv:hep-ph/0701166} \BibitemShut {NoStop}%
\bibitem [{\citenamefont {Dimopoulos}(1980)}]{Dimopoulos:1979sp}%
  \BibitemOpen
  \bibfield  {author} {\bibinfo {author} {\bibfnamefont {S.}~\bibnamefont
  {Dimopoulos}},\ }\bibfield  {title} {\bibinfo {title} {{Technicolored
  Signatures}},\ }\href {https://doi.org/10.1016/0550-3213(80)90277-1}
  {\bibfield  {journal} {\bibinfo  {journal} {Nucl. Phys. B}\ }\textbf
  {\bibinfo {volume} {168}},\ \bibinfo {pages} {69} (\bibinfo {year}
  {1980})}\BibitemShut {NoStop}%
\bibitem [{\citenamefont {Farhi}\ and\ \citenamefont
  {Susskind}(1981)}]{Farhi:1980xs}%
  \BibitemOpen
  \bibfield  {author} {\bibinfo {author} {\bibfnamefont {E.}~\bibnamefont
  {Farhi}}\ and\ \bibinfo {author} {\bibfnamefont {L.}~\bibnamefont
  {Susskind}},\ }\bibfield  {title} {\bibinfo {title} {{Technicolor}},\ }\href
  {https://doi.org/10.1016/0370-1573(81)90173-3} {\bibfield  {journal}
  {\bibinfo  {journal} {Phys. Rept.}\ }\textbf {\bibinfo {volume} {74}},\
  \bibinfo {pages} {277} (\bibinfo {year} {1981})}\BibitemShut {NoStop}%
\bibitem [{\citenamefont {Hill}\ \emph {et~al.}(1993)\citenamefont {Hill},
  \citenamefont {Kennedy}, \citenamefont {Onogi},\ and\ \citenamefont
  {Yu}}]{Hill:1992ev}%
  \BibitemOpen
  \bibfield  {author} {\bibinfo {author} {\bibfnamefont {C.~T.}\ \bibnamefont
  {Hill}}, \bibinfo {author} {\bibfnamefont {D.~C.}\ \bibnamefont {Kennedy}},
  \bibinfo {author} {\bibfnamefont {T.}~\bibnamefont {Onogi}},\ and\ \bibinfo
  {author} {\bibfnamefont {H.-L.}\ \bibnamefont {Yu}},\ }\bibfield  {title}
  {\bibinfo {title} {{Spontaneously broken technicolor and the dynamics of
  virtual vector technimesons}},\ }\href
  {https://doi.org/10.1103/PhysRevD.47.2940} {\bibfield  {journal} {\bibinfo
  {journal} {Phys. Rev. D}\ }\textbf {\bibinfo {volume} {47}},\ \bibinfo
  {pages} {2940} (\bibinfo {year} {1993})},\ \Eprint
  {https://arxiv.org/abs/hep-ph/9210233} {arXiv:hep-ph/9210233} \BibitemShut
  {NoStop}%
\bibitem [{ATL(2019)}]{ATLAS:2019vcr}%
  \BibitemOpen
  \href@noop {} {\bibinfo {title} {{Search for high-mass dilepton resonances
  using $139\,\mathrm{fb}^{-1}$ of $pp$ collision data collected at
  $\sqrt{s}=13\,\mathrm{TeV}$ with the ATLAS detector}}} (\bibinfo {year}
  {2019})\BibitemShut {NoStop}%
\bibitem [{\citenamefont {Ball}\ \emph
  {et~al.}(2013{\natexlab{b}})\citenamefont {Ball}, \citenamefont {Bertone},
  \citenamefont {Carrazza}, \citenamefont {Del~Debbio}, \citenamefont {Forte},
  \citenamefont {Guffanti}, \citenamefont {Hartland},\ and\ \citenamefont
  {Rojo}}]{Ball:2013hta}%
  \BibitemOpen
  \bibfield  {author} {\bibinfo {author} {\bibfnamefont {R.~D.}\ \bibnamefont
  {Ball}}, \bibinfo {author} {\bibfnamefont {V.}~\bibnamefont {Bertone}},
  \bibinfo {author} {\bibfnamefont {S.}~\bibnamefont {Carrazza}}, \bibinfo
  {author} {\bibfnamefont {L.}~\bibnamefont {Del~Debbio}}, \bibinfo {author}
  {\bibfnamefont {S.}~\bibnamefont {Forte}}, \bibinfo {author} {\bibfnamefont
  {A.}~\bibnamefont {Guffanti}}, \bibinfo {author} {\bibfnamefont {N.~P.}\
  \bibnamefont {Hartland}},\ and\ \bibinfo {author} {\bibfnamefont
  {J.}~\bibnamefont {Rojo}} (\bibinfo {collaboration} {NNPDF}),\ }\bibfield
  {title} {\bibinfo {title} {{Parton distributions with QED corrections}},\
  }\href {https://doi.org/10.1016/j.nuclphysb.2013.10.010} {\bibfield
  {journal} {\bibinfo  {journal} {Nucl. Phys. B}\ }\textbf {\bibinfo {volume}
  {877}},\ \bibinfo {pages} {290} (\bibinfo {year} {2013}{\natexlab{b}})},\
  \Eprint {https://arxiv.org/abs/1308.0598} {arXiv:1308.0598 [hep-ph]}
  \BibitemShut {NoStop}%
\bibitem [{\citenamefont {Clark}\ \emph {et~al.}(2017)\citenamefont {Clark},
  \citenamefont {Godat},\ and\ \citenamefont {Olness}}]{Clark:2016jgm}%
  \BibitemOpen
  \bibfield  {author} {\bibinfo {author} {\bibfnamefont {D.~B.}\ \bibnamefont
  {Clark}}, \bibinfo {author} {\bibfnamefont {E.}~\bibnamefont {Godat}},\ and\
  \bibinfo {author} {\bibfnamefont {F.~I.}\ \bibnamefont {Olness}},\ }\bibfield
   {title} {\bibinfo {title} {{ManeParse: A Mathematica reader for Parton
  Distribution Functions}},\ }\href {https://doi.org/10.1016/j.cpc.2017.03.004}
  {\bibfield  {journal} {\bibinfo  {journal} {Comput. Phys. Commun.}\ }\textbf
  {\bibinfo {volume} {216}},\ \bibinfo {pages} {126} (\bibinfo {year}
  {2017})},\ \Eprint {https://arxiv.org/abs/1605.08012} {arXiv:1605.08012
  [hep-ph]} \BibitemShut {NoStop}%
\bibitem [{\citenamefont {de~Florian}\ \emph {et~al.}(2016)\citenamefont
  {de~Florian} \emph {et~al.}}]{LHCHiggsCrossSectionWorkingGroup:2016ypw}%
  \BibitemOpen
  \bibfield  {author} {\bibinfo {author} {\bibfnamefont {D.}~\bibnamefont
  {de~Florian}} \emph {et~al.} (\bibinfo {collaboration} {LHC Higgs Cross
  Section Working Group}),\ }\href {https://doi.org/10.23731/CYRM-2017-002}
  {\bibinfo {title} {{Handbook of LHC Higgs Cross Sections: 4. Deciphering the
  Nature of the Higgs Sector}}} (\bibinfo {year} {2016}),\ \Eprint
  {https://arxiv.org/abs/1610.07922} {arXiv:1610.07922 [hep-ph]} \BibitemShut
  {NoStop}%
\bibitem [{\citenamefont {Cakir}\ and\ \citenamefont
  {Sahin}(2005)}]{Cakir:2005iw}%
  \BibitemOpen
  \bibfield  {author} {\bibinfo {author} {\bibfnamefont {O.}~\bibnamefont
  {Cakir}}\ and\ \bibinfo {author} {\bibfnamefont {M.}~\bibnamefont {Sahin}},\
  }\bibfield  {title} {\bibinfo {title} {{Resonant production of diquarks at
  high energy $p p$, ep and $e^{+} e^{-}$ colliders}},\ }\href
  {https://doi.org/10.1103/PhysRevD.72.115011} {\bibfield  {journal} {\bibinfo
  {journal} {Phys. Rev. D}\ }\textbf {\bibinfo {volume} {72}},\ \bibinfo
  {pages} {115011} (\bibinfo {year} {2005})},\ \Eprint
  {https://arxiv.org/abs/hep-ph/0508205} {arXiv:hep-ph/0508205} \BibitemShut
  {NoStop}%
\bibitem [{\citenamefont {Mohapatra}\ \emph {et~al.}(2008)\citenamefont
  {Mohapatra}, \citenamefont {Okada},\ and\ \citenamefont
  {Yu}}]{Mohapatra:2007af}%
  \BibitemOpen
  \bibfield  {author} {\bibinfo {author} {\bibfnamefont {R.~N.}\ \bibnamefont
  {Mohapatra}}, \bibinfo {author} {\bibfnamefont {N.}~\bibnamefont {Okada}},\
  and\ \bibinfo {author} {\bibfnamefont {H.-B.}\ \bibnamefont {Yu}},\
  }\bibfield  {title} {\bibinfo {title} {{Diquark Higgs at LHC}},\ }\href
  {https://doi.org/10.1103/PhysRevD.77.011701} {\bibfield  {journal} {\bibinfo
  {journal} {Phys. Rev. D}\ }\textbf {\bibinfo {volume} {77}},\ \bibinfo
  {pages} {011701} (\bibinfo {year} {2008})},\ \Eprint
  {https://arxiv.org/abs/0709.1486} {arXiv:0709.1486 [hep-ph]} \BibitemShut
  {NoStop}%
\bibitem [{\citenamefont {Chen}\ \emph {et~al.}(2009)\citenamefont {Chen},
  \citenamefont {Klemm}, \citenamefont {Rentala},\ and\ \citenamefont
  {Wang}}]{Chen:2008hh}%
  \BibitemOpen
  \bibfield  {author} {\bibinfo {author} {\bibfnamefont {C.-R.}\ \bibnamefont
  {Chen}}, \bibinfo {author} {\bibfnamefont {W.}~\bibnamefont {Klemm}},
  \bibinfo {author} {\bibfnamefont {V.}~\bibnamefont {Rentala}},\ and\ \bibinfo
  {author} {\bibfnamefont {K.}~\bibnamefont {Wang}},\ }\bibfield  {title}
  {\bibinfo {title} {{Color Sextet Scalars at the CERN Large Hadron
  Collider}},\ }\href {https://doi.org/10.1103/PhysRevD.79.054002} {\bibfield
  {journal} {\bibinfo  {journal} {Phys. Rev. D}\ }\textbf {\bibinfo {volume}
  {79}},\ \bibinfo {pages} {054002} (\bibinfo {year} {2009})},\ \Eprint
  {https://arxiv.org/abs/0811.2105} {arXiv:0811.2105 [hep-ph]} \BibitemShut
  {NoStop}%
\bibitem [{\citenamefont {Kilic}\ \emph {et~al.}(2008)\citenamefont {Kilic},
  \citenamefont {Okui},\ and\ \citenamefont {Sundrum}}]{Kilic:2008pm}%
  \BibitemOpen
  \bibfield  {author} {\bibinfo {author} {\bibfnamefont {C.}~\bibnamefont
  {Kilic}}, \bibinfo {author} {\bibfnamefont {T.}~\bibnamefont {Okui}},\ and\
  \bibinfo {author} {\bibfnamefont {R.}~\bibnamefont {Sundrum}},\ }\bibfield
  {title} {\bibinfo {title} {{Colored Resonances at the Tevatron: Phenomenology
  and Discovery Potential in Multijets}},\ }\href
  {https://doi.org/10.1088/1126-6708/2008/07/038} {\bibfield  {journal}
  {\bibinfo  {journal} {JHEP}\ }\textbf {\bibinfo {volume} {07}},\ \bibinfo
  {pages} {038}},\ \Eprint {https://arxiv.org/abs/0802.2568} {arXiv:0802.2568
  [hep-ph]} \BibitemShut {NoStop}%
\bibitem [{\citenamefont {Berger}\ \emph {et~al.}(2010)\citenamefont {Berger},
  \citenamefont {Cao}, \citenamefont {Chen}, \citenamefont {Shaughnessy},\ and\
  \citenamefont {Zhang}}]{Berger:2010fy}%
  \BibitemOpen
  \bibfield  {author} {\bibinfo {author} {\bibfnamefont {E.~L.}\ \bibnamefont
  {Berger}}, \bibinfo {author} {\bibfnamefont {Q.-H.}\ \bibnamefont {Cao}},
  \bibinfo {author} {\bibfnamefont {C.-R.}\ \bibnamefont {Chen}}, \bibinfo
  {author} {\bibfnamefont {G.}~\bibnamefont {Shaughnessy}},\ and\ \bibinfo
  {author} {\bibfnamefont {H.}~\bibnamefont {Zhang}},\ }\bibfield  {title}
  {\bibinfo {title} {{Color Sextet Scalars in Early LHC Experiments}},\ }\href
  {https://doi.org/10.1103/PhysRevLett.105.181802} {\bibfield  {journal}
  {\bibinfo  {journal} {Phys. Rev. Lett.}\ }\textbf {\bibinfo {volume} {105}},\
  \bibinfo {pages} {181802} (\bibinfo {year} {2010})},\ \Eprint
  {https://arxiv.org/abs/1005.2622} {arXiv:1005.2622 [hep-ph]} \BibitemShut
  {NoStop}%
\bibitem [{\citenamefont {Bai}\ and\ \citenamefont
  {Shelton}(2012)}]{Bai:2011mr}%
  \BibitemOpen
  \bibfield  {author} {\bibinfo {author} {\bibfnamefont {Y.}~\bibnamefont
  {Bai}}\ and\ \bibinfo {author} {\bibfnamefont {J.}~\bibnamefont {Shelton}},\
  }\bibfield  {title} {\bibinfo {title} {{Composite Octet Searches with Jet
  Substructure}},\ }\href {https://doi.org/10.1007/JHEP07(2012)067} {\bibfield
  {journal} {\bibinfo  {journal} {JHEP}\ }\textbf {\bibinfo {volume} {07}},\
  \bibinfo {pages} {067}},\ \Eprint {https://arxiv.org/abs/1107.3563}
  {arXiv:1107.3563 [hep-ph]} \BibitemShut {NoStop}%
\bibitem [{\citenamefont {Dobrescu}\ \emph {et~al.}(2018)\citenamefont
  {Dobrescu}, \citenamefont {Harris},\ and\ \citenamefont
  {Isaacson}}]{Dobrescu:2018psr}%
  \BibitemOpen
  \bibfield  {author} {\bibinfo {author} {\bibfnamefont {B.~A.}\ \bibnamefont
  {Dobrescu}}, \bibinfo {author} {\bibfnamefont {R.~M.}\ \bibnamefont
  {Harris}},\ and\ \bibinfo {author} {\bibfnamefont {J.}~\bibnamefont
  {Isaacson}},\ }\href@noop {} {\bibinfo {title} {{Ultraheavy resonances at the
  LHC: beyond the QCD background}}} (\bibinfo {year} {2018}),\ \Eprint
  {https://arxiv.org/abs/1810.09429} {arXiv:1810.09429 [hep-ph]} \BibitemShut
  {NoStop}%
\bibitem [{\citenamefont {Pascual-Dias}\ \emph {et~al.}(2020)\citenamefont
  {Pascual-Dias}, \citenamefont {Saha},\ and\ \citenamefont
  {London}}]{Pascual-Dias:2020hxo}%
  \BibitemOpen
  \bibfield  {author} {\bibinfo {author} {\bibfnamefont {B.}~\bibnamefont
  {Pascual-Dias}}, \bibinfo {author} {\bibfnamefont {P.}~\bibnamefont {Saha}},\
  and\ \bibinfo {author} {\bibfnamefont {D.}~\bibnamefont {London}},\
  }\bibfield  {title} {\bibinfo {title} {{LHC Constraints on Scalar
  Diquarks}},\ }\href {https://doi.org/10.1007/JHEP07(2020)144} {\bibfield
  {journal} {\bibinfo  {journal} {JHEP}\ }\textbf {\bibinfo {volume} {07}},\
  \bibinfo {pages} {144}},\ \Eprint {https://arxiv.org/abs/2006.13385}
  {arXiv:2006.13385 [hep-ph]} \BibitemShut {NoStop}%
\bibitem [{\citenamefont {Baldo-Ceolin}\ \emph {et~al.}(1994)\citenamefont
  {Baldo-Ceolin} \emph {et~al.}}]{Baldo-Ceolin:1994hzw}%
  \BibitemOpen
  \bibfield  {author} {\bibinfo {author} {\bibfnamefont {M.}~\bibnamefont
  {Baldo-Ceolin}} \emph {et~al.},\ }\bibfield  {title} {\bibinfo {title} {{A
  New experimental limit on neutron - anti-neutron oscillations}},\ }\href
  {https://doi.org/10.1007/BF01580321} {\bibfield  {journal} {\bibinfo
  {journal} {Z. Phys. C}\ }\textbf {\bibinfo {volume} {63}},\ \bibinfo {pages}
  {409} (\bibinfo {year} {1994})}\BibitemShut {NoStop}%
\bibitem [{\citenamefont {Randall}\ and\ \citenamefont
  {Sundrum}(1999)}]{Randall:1999ee}%
  \BibitemOpen
  \bibfield  {author} {\bibinfo {author} {\bibfnamefont {L.}~\bibnamefont
  {Randall}}\ and\ \bibinfo {author} {\bibfnamefont {R.}~\bibnamefont
  {Sundrum}},\ }\bibfield  {title} {\bibinfo {title} {{A Large mass hierarchy
  from a small extra dimension}},\ }\href
  {https://doi.org/10.1103/PhysRevLett.83.3370} {\bibfield  {journal} {\bibinfo
   {journal} {Phys. Rev. Lett.}\ }\textbf {\bibinfo {volume} {83}},\ \bibinfo
  {pages} {3370} (\bibinfo {year} {1999})},\ \Eprint
  {https://arxiv.org/abs/hep-ph/9905221} {arXiv:hep-ph/9905221} \BibitemShut
  {NoStop}%
\bibitem [{\citenamefont {Gherghetta}\ and\ \citenamefont
  {Pomarol}(2000)}]{Gherghetta:2000qt}%
  \BibitemOpen
  \bibfield  {author} {\bibinfo {author} {\bibfnamefont {T.}~\bibnamefont
  {Gherghetta}}\ and\ \bibinfo {author} {\bibfnamefont {A.}~\bibnamefont
  {Pomarol}},\ }\bibfield  {title} {\bibinfo {title} {{Bulk fields and
  supersymmetry in a slice of AdS}},\ }\href
  {https://doi.org/10.1016/S0550-3213(00)00392-8} {\bibfield  {journal}
  {\bibinfo  {journal} {Nucl. Phys. B}\ }\textbf {\bibinfo {volume} {586}},\
  \bibinfo {pages} {141} (\bibinfo {year} {2000})},\ \Eprint
  {https://arxiv.org/abs/hep-ph/0003129} {arXiv:hep-ph/0003129} \BibitemShut
  {NoStop}%
\bibitem [{\citenamefont {Aaij}\ \emph {et~al.}(2022)\citenamefont {Aaij} \emph
  {et~al.}}]{LHCb:2021trn}%
  \BibitemOpen
  \bibfield  {author} {\bibinfo {author} {\bibfnamefont {R.}~\bibnamefont
  {Aaij}} \emph {et~al.} (\bibinfo {collaboration} {LHCb}),\ }\bibfield
  {title} {\bibinfo {title} {{Test of lepton universality in beauty-quark
  decays}},\ }\href {https://doi.org/10.1038/s41567-021-01478-8} {\bibfield
  {journal} {\bibinfo  {journal} {Nature Phys.}\ }\textbf {\bibinfo {volume}
  {18}},\ \bibinfo {pages} {277} (\bibinfo {year} {2022})},\ \Eprint
  {https://arxiv.org/abs/2103.11769} {arXiv:2103.11769 [hep-ex]} \BibitemShut
  {NoStop}%
\bibitem [{\citenamefont {Amhis}\ \emph {et~al.}(2021)\citenamefont {Amhis}
  \emph {et~al.}}]{HFLAV:2019otj}%
  \BibitemOpen
  \bibfield  {author} {\bibinfo {author} {\bibfnamefont {Y.~S.}\ \bibnamefont
  {Amhis}} \emph {et~al.} (\bibinfo {collaboration} {HFLAV}),\ }\bibfield
  {title} {\bibinfo {title} {{Averages of b-hadron, c-hadron, and $\tau
  $-lepton properties as of 2018}},\ }\href
  {https://doi.org/10.1140/epjc/s10052-020-8156-7} {\bibfield  {journal}
  {\bibinfo  {journal} {Eur. Phys. J. C}\ }\textbf {\bibinfo {volume} {81}},\
  \bibinfo {pages} {226} (\bibinfo {year} {2021})},\ \Eprint
  {https://arxiv.org/abs/1909.12524} {arXiv:1909.12524 [hep-ex]} \BibitemShut
  {NoStop}%
\bibitem [{\citenamefont {Bennett}\ \emph {et~al.}(2006)\citenamefont {Bennett}
  \emph {et~al.}}]{Bennett:2006fi}%
  \BibitemOpen
  \bibfield  {author} {\bibinfo {author} {\bibfnamefont {G.~W.}\ \bibnamefont
  {Bennett}} \emph {et~al.} (\bibinfo {collaboration} {Muon g-2}),\ }\bibfield
  {title} {\bibinfo {title} {{Final Report of the Muon E821 Anomalous Magnetic
  Moment Measurement at BNL}},\ }\href
  {https://doi.org/10.1103/PhysRevD.73.072003} {\bibfield  {journal} {\bibinfo
  {journal} {Phys. Rev. D}\ }\textbf {\bibinfo {volume} {73}},\ \bibinfo
  {pages} {072003} (\bibinfo {year} {2006})},\ \Eprint
  {https://arxiv.org/abs/hep-ex/0602035} {arXiv:hep-ex/0602035} \BibitemShut
  {NoStop}%
\bibitem [{\citenamefont {Abi}\ \emph {et~al.}(2021)\citenamefont {Abi} \emph
  {et~al.}}]{Abi:2021gix}%
  \BibitemOpen
  \bibfield  {author} {\bibinfo {author} {\bibfnamefont {B.}~\bibnamefont
  {Abi}} \emph {et~al.} (\bibinfo {collaboration} {Muon g-2}),\ }\bibfield
  {title} {\bibinfo {title} {{Measurement of the Positive Muon Anomalous
  Magnetic Moment to 0.46 ppm}},\ }\href
  {https://doi.org/10.1103/PhysRevLett.126.141801} {\bibfield  {journal}
  {\bibinfo  {journal} {Phys. Rev. Lett.}\ }\textbf {\bibinfo {volume} {126}},\
  \bibinfo {pages} {141801} (\bibinfo {year} {2021})},\ \Eprint
  {https://arxiv.org/abs/2104.03281} {arXiv:2104.03281 [hep-ex]} \BibitemShut
  {NoStop}%
\bibitem [{\citenamefont {Aoyama}\ \emph {et~al.}(2019)\citenamefont {Aoyama},
  \citenamefont {Kinoshita},\ and\ \citenamefont {Nio}}]{Aoyama:2019ryr}%
  \BibitemOpen
  \bibfield  {author} {\bibinfo {author} {\bibfnamefont {T.}~\bibnamefont
  {Aoyama}}, \bibinfo {author} {\bibfnamefont {T.}~\bibnamefont {Kinoshita}},\
  and\ \bibinfo {author} {\bibfnamefont {M.}~\bibnamefont {Nio}},\ }\bibfield
  {title} {\bibinfo {title} {{Theory of the Anomalous Magnetic Moment of the
  Electron}},\ }\href {https://doi.org/10.3390/atoms7010028} {\bibfield
  {journal} {\bibinfo  {journal} {Atoms}\ }\textbf {\bibinfo {volume} {7}},\
  \bibinfo {pages} {28} (\bibinfo {year} {2019})}\BibitemShut {NoStop}%
\bibitem [{\citenamefont {Aaltonen}\ \emph {et~al.}(2022)\citenamefont
  {Aaltonen} \emph {et~al.}}]{CDF:2022hxs}%
  \BibitemOpen
  \bibfield  {author} {\bibinfo {author} {\bibfnamefont {T.}~\bibnamefont
  {Aaltonen}} \emph {et~al.} (\bibinfo {collaboration} {CDF}),\ }\bibfield
  {title} {\bibinfo {title} {{High-precision measurement of the W boson mass
  with the CDF II detector}},\ }\href {https://doi.org/10.1126/science.abk1781}
  {\bibfield  {journal} {\bibinfo  {journal} {Science}\ }\textbf {\bibinfo
  {volume} {376}},\ \bibinfo {pages} {170} (\bibinfo {year}
  {2022})}\BibitemShut {NoStop}%
\bibitem [{\citenamefont {Zyla}\ \emph {et~al.}(2020)\citenamefont {Zyla} \emph
  {et~al.}}]{ParticleDataGroup:2020ssz}%
  \BibitemOpen
  \bibfield  {author} {\bibinfo {author} {\bibfnamefont {P.~A.}\ \bibnamefont
  {Zyla}} \emph {et~al.} (\bibinfo {collaboration} {Particle Data Group}),\
  }\bibfield  {title} {\bibinfo {title} {{Review of Particle Physics}},\ }\href
  {https://doi.org/10.1093/ptep/ptaa104} {\bibfield  {journal} {\bibinfo
  {journal} {PTEP}\ }\textbf {\bibinfo {volume} {2020}},\ \bibinfo {pages}
  {083C01} (\bibinfo {year} {2020})}\BibitemShut {NoStop}%
\bibitem [{\citenamefont {Seng}(2022)}]{Seng:2022ufm}%
  \BibitemOpen
  \bibfield  {author} {\bibinfo {author} {\bibfnamefont {C.-Y.}\ \bibnamefont
  {Seng}},\ }\bibfield  {title} {\bibinfo {title} {{First row CKM unitarity}},\
  }in\ \href@noop {} {\emph {\bibinfo {booktitle} {{20th Conference on Flavor
  Physics and CP Violation~}}}}\ (\bibinfo {year} {2022})\ \Eprint
  {https://arxiv.org/abs/2207.10492} {arXiv:2207.10492 [hep-ph]} \BibitemShut
  {NoStop}%
\bibitem [{\citenamefont {Manzari}(2022)}]{Manzari:2021kma}%
  \BibitemOpen
  \bibfield  {author} {\bibinfo {author} {\bibfnamefont {C.~A.}\ \bibnamefont
  {Manzari}},\ }\bibfield  {title} {\bibinfo {title} {{Explaining the Cabibbo
  Angle Anomaly}},\ }\href {https://doi.org/10.22323/1.398.0526} {\bibfield
  {journal} {\bibinfo  {journal} {PoS}\ }\textbf {\bibinfo {volume}
  {EPS-HEP2021}},\ \bibinfo {pages} {526} (\bibinfo {year} {2022})},\ \Eprint
  {https://arxiv.org/abs/2111.04519} {arXiv:2111.04519 [hep-ph]} \BibitemShut
  {NoStop}%
\bibitem [{\citenamefont {Crivellin}(2022)}]{Crivellin:2022ctt}%
  \BibitemOpen
  \bibfield  {author} {\bibinfo {author} {\bibfnamefont {A.}~\bibnamefont
  {Crivellin}},\ }\bibfield  {title} {\bibinfo {title} {{Explaining the Cabibbo
  Angle Anomaly}}\ }(\bibinfo {year} {2022})\ \Eprint
  {https://arxiv.org/abs/2207.02507} {arXiv:2207.02507 [hep-ph]} \BibitemShut
  {NoStop}%
\bibitem [{CMS(2017)}]{CMS-PAS-HIG-17-013}%
  \BibitemOpen
  \href@noop {} {\bibinfo {title} {{Search for new resonances in the diphoton
  final state in the mass range between 70 and 110 GeV in pp collisions at
  $\sqrt{s}=$ 8 and 13 TeV}}} (\bibinfo {year} {2017})\BibitemShut {NoStop}%
\bibitem [{\citenamefont {Crivellin}\ \emph
  {et~al.}(2021{\natexlab{a}})\citenamefont {Crivellin}, \citenamefont {Fang},
  \citenamefont {Fischer}, \citenamefont {Kumar}, \citenamefont {Kumar},
  \citenamefont {Malwa}, \citenamefont {Mellado}, \citenamefont {Rapheeha},
  \citenamefont {Ruan},\ and\ \citenamefont {Sha}}]{Crivellin:2021ubm}%
  \BibitemOpen
  \bibfield  {author} {\bibinfo {author} {\bibfnamefont {A.}~\bibnamefont
  {Crivellin}}, \bibinfo {author} {\bibfnamefont {Y.}~\bibnamefont {Fang}},
  \bibinfo {author} {\bibfnamefont {O.}~\bibnamefont {Fischer}}, \bibinfo
  {author} {\bibfnamefont {A.}~\bibnamefont {Kumar}}, \bibinfo {author}
  {\bibfnamefont {M.}~\bibnamefont {Kumar}}, \bibinfo {author} {\bibfnamefont
  {E.}~\bibnamefont {Malwa}}, \bibinfo {author} {\bibfnamefont
  {B.}~\bibnamefont {Mellado}}, \bibinfo {author} {\bibfnamefont
  {N.}~\bibnamefont {Rapheeha}}, \bibinfo {author} {\bibfnamefont
  {X.}~\bibnamefont {Ruan}},\ and\ \bibinfo {author} {\bibfnamefont
  {Q.}~\bibnamefont {Sha}},\ }\href@noop {} {\bibinfo {title} {{Accumulating
  Evidence for the Associate Production of a Neutral Scalar with Mass around
  151 GeV}}} (\bibinfo {year} {2021}{\natexlab{a}}),\ \Eprint
  {https://arxiv.org/abs/2109.02650} {arXiv:2109.02650 [hep-ph]} \BibitemShut
  {NoStop}%
\bibitem [{\citenamefont {Aaboud}\ \emph {et~al.}(2017)\citenamefont {Aaboud}
  \emph {et~al.}}]{ATLAS:2017ayi}%
  \BibitemOpen
  \bibfield  {author} {\bibinfo {author} {\bibfnamefont {M.}~\bibnamefont
  {Aaboud}} \emph {et~al.} (\bibinfo {collaboration} {ATLAS}),\ }\bibfield
  {title} {\bibinfo {title} {{Search for new phenomena in high-mass diphoton
  final states using 37 fb$^{-1}$ of proton--proton collisions collected at
  $\sqrt{s}=13$ TeV with the ATLAS detector}},\ }\href
  {https://doi.org/10.1016/j.physletb.2017.10.039} {\bibfield  {journal}
  {\bibinfo  {journal} {Phys. Lett. B}\ }\textbf {\bibinfo {volume} {775}},\
  \bibinfo {pages} {105} (\bibinfo {year} {2017})},\ \Eprint
  {https://arxiv.org/abs/1707.04147} {arXiv:1707.04147 [hep-ex]} \BibitemShut
  {NoStop}%
\bibitem [{\citenamefont {von Buddenbrock}\ \emph {et~al.}(2016)\citenamefont
  {von Buddenbrock}, \citenamefont {Chakrabarty}, \citenamefont {Cornell},
  \citenamefont {Kar}, \citenamefont {Kumar}, \citenamefont {Mandal},
  \citenamefont {Mellado}, \citenamefont {Mukhopadhyaya}, \citenamefont
  {Reed},\ and\ \citenamefont {Ruan}}]{vonBuddenbrock:2016rmr}%
  \BibitemOpen
  \bibfield  {author} {\bibinfo {author} {\bibfnamefont {S.}~\bibnamefont {von
  Buddenbrock}}, \bibinfo {author} {\bibfnamefont {N.}~\bibnamefont
  {Chakrabarty}}, \bibinfo {author} {\bibfnamefont {A.~S.}\ \bibnamefont
  {Cornell}}, \bibinfo {author} {\bibfnamefont {D.}~\bibnamefont {Kar}},
  \bibinfo {author} {\bibfnamefont {M.}~\bibnamefont {Kumar}}, \bibinfo
  {author} {\bibfnamefont {T.}~\bibnamefont {Mandal}}, \bibinfo {author}
  {\bibfnamefont {B.}~\bibnamefont {Mellado}}, \bibinfo {author} {\bibfnamefont
  {B.}~\bibnamefont {Mukhopadhyaya}}, \bibinfo {author} {\bibfnamefont {R.~G.}\
  \bibnamefont {Reed}},\ and\ \bibinfo {author} {\bibfnamefont
  {X.}~\bibnamefont {Ruan}},\ }\bibfield  {title} {\bibinfo {title}
  {{Phenomenological signatures of additional scalar bosons at the LHC}},\
  }\href {https://doi.org/10.1140/epjc/s10052-016-4435-8} {\bibfield  {journal}
  {\bibinfo  {journal} {Eur. Phys. J. C}\ }\textbf {\bibinfo {volume} {76}},\
  \bibinfo {pages} {580} (\bibinfo {year} {2016})},\ \Eprint
  {https://arxiv.org/abs/1606.01674} {arXiv:1606.01674 [hep-ph]} \BibitemShut
  {NoStop}%
\bibitem [{\citenamefont {Buddenbrock}\ \emph {et~al.}(2019)\citenamefont
  {Buddenbrock}, \citenamefont {Cornell}, \citenamefont {Fang}, \citenamefont
  {Fadol~Mohammed}, \citenamefont {Kumar}, \citenamefont {Mellado},\ and\
  \citenamefont {Tomiwa}}]{vonBuddenbrock:2019ajh}%
  \BibitemOpen
  \bibfield  {author} {\bibinfo {author} {\bibfnamefont {S.}~\bibnamefont
  {Buddenbrock}}, \bibinfo {author} {\bibfnamefont {A.~S.}\ \bibnamefont
  {Cornell}}, \bibinfo {author} {\bibfnamefont {Y.}~\bibnamefont {Fang}},
  \bibinfo {author} {\bibfnamefont {A.}~\bibnamefont {Fadol~Mohammed}},
  \bibinfo {author} {\bibfnamefont {M.}~\bibnamefont {Kumar}}, \bibinfo
  {author} {\bibfnamefont {B.}~\bibnamefont {Mellado}},\ and\ \bibinfo {author}
  {\bibfnamefont {K.~G.}\ \bibnamefont {Tomiwa}},\ }\bibfield  {title}
  {\bibinfo {title} {{The emergence of multi-lepton anomalies at the LHC and
  their compatibility with new physics at the EW scale}},\ }\href
  {https://doi.org/10.1007/JHEP10(2019)157} {\bibfield  {journal} {\bibinfo
  {journal} {JHEP}\ }\textbf {\bibinfo {volume} {10}},\ \bibinfo {pages}
  {157}},\ \Eprint {https://arxiv.org/abs/1901.05300} {arXiv:1901.05300
  [hep-ph]} \BibitemShut {NoStop}%
\bibitem [{\citenamefont {von Buddenbrock}\ \emph {et~al.}(2020)\citenamefont
  {von Buddenbrock}, \citenamefont {Ruiz},\ and\ \citenamefont
  {Mellado}}]{vonBuddenbrock:2020ter}%
  \BibitemOpen
  \bibfield  {author} {\bibinfo {author} {\bibfnamefont {S.}~\bibnamefont {von
  Buddenbrock}}, \bibinfo {author} {\bibfnamefont {R.}~\bibnamefont {Ruiz}},\
  and\ \bibinfo {author} {\bibfnamefont {B.}~\bibnamefont {Mellado}},\
  }\bibfield  {title} {\bibinfo {title} {{Anatomy of inclusive $t\bar t W$
  production at hadron colliders}},\ }\href
  {https://doi.org/10.1016/j.physletb.2020.135964} {\bibfield  {journal}
  {\bibinfo  {journal} {Phys. Lett. B}\ }\textbf {\bibinfo {volume} {811}},\
  \bibinfo {pages} {135964} (\bibinfo {year} {2020})},\ \Eprint
  {https://arxiv.org/abs/2009.00032} {arXiv:2009.00032 [hep-ph]} \BibitemShut
  {NoStop}%
\bibitem [{\citenamefont {Hernandez}\ \emph {et~al.}(2021)\citenamefont
  {Hernandez}, \citenamefont {Kumar}, \citenamefont {Cornell}, \citenamefont
  {Dahbi}, \citenamefont {Fang}, \citenamefont {Lieberman}, \citenamefont
  {Mellado}, \citenamefont {Monnakgotla}, \citenamefont {Ruan},\ and\
  \citenamefont {Xin}}]{Hernandez:2019geu}%
  \BibitemOpen
  \bibfield  {author} {\bibinfo {author} {\bibfnamefont {Y.}~\bibnamefont
  {Hernandez}}, \bibinfo {author} {\bibfnamefont {M.}~\bibnamefont {Kumar}},
  \bibinfo {author} {\bibfnamefont {A.~S.}\ \bibnamefont {Cornell}}, \bibinfo
  {author} {\bibfnamefont {S.-E.}\ \bibnamefont {Dahbi}}, \bibinfo {author}
  {\bibfnamefont {Y.}~\bibnamefont {Fang}}, \bibinfo {author} {\bibfnamefont
  {B.}~\bibnamefont {Lieberman}}, \bibinfo {author} {\bibfnamefont
  {B.}~\bibnamefont {Mellado}}, \bibinfo {author} {\bibfnamefont
  {K.}~\bibnamefont {Monnakgotla}}, \bibinfo {author} {\bibfnamefont
  {X.}~\bibnamefont {Ruan}},\ and\ \bibinfo {author} {\bibfnamefont
  {S.}~\bibnamefont {Xin}},\ }\bibfield  {title} {\bibinfo {title} {{The
  anomalous production of multi-lepton and its impact on the measurement of
  $Wh$ production at the LHC}},\ }\href
  {https://doi.org/10.1140/epjc/s10052-021-09137-1} {\bibfield  {journal}
  {\bibinfo  {journal} {Eur. Phys. J. C}\ }\textbf {\bibinfo {volume} {81}},\
  \bibinfo {pages} {365} (\bibinfo {year} {2021})},\ \Eprint
  {https://arxiv.org/abs/1912.00699} {arXiv:1912.00699 [hep-ph]} \BibitemShut
  {NoStop}%
\bibitem [{ATL(2021)}]{ATLAS:2021tyg}%
  \BibitemOpen
  \href@noop {} {\bibinfo {title} {{Combination of searches for non-resonant
  and resonant Higgs boson pair production in the $b\bar{b}\gamma\gamma$,
  $b\bar{b}\tau^{+}\tau^{-}$ and $b\bar{b}b\bar{b}$ decay channels using $pp$
  collisions at $\sqrt{s}$ = 13 TeV with the ATLAS detector}}} (\bibinfo {year}
  {2021})\BibitemShut {NoStop}%
\bibitem [{\citenamefont {Crivellin}\ \emph
  {et~al.}(2021{\natexlab{b}})\citenamefont {Crivellin}, \citenamefont
  {Manzari},\ and\ \citenamefont {Montull}}]{Crivellin:2021rbf}%
  \BibitemOpen
  \bibfield  {author} {\bibinfo {author} {\bibfnamefont {A.}~\bibnamefont
  {Crivellin}}, \bibinfo {author} {\bibfnamefont {C.~A.}\ \bibnamefont
  {Manzari}},\ and\ \bibinfo {author} {\bibfnamefont {M.}~\bibnamefont
  {Montull}},\ }\bibfield  {title} {\bibinfo {title} {{Correlating nonresonant
  di-electron searches at the LHC to the Cabibbo-angle anomaly and lepton
  flavor universality violation}},\ }\href
  {https://doi.org/10.1103/PhysRevD.104.115016} {\bibfield  {journal} {\bibinfo
   {journal} {Phys. Rev. D}\ }\textbf {\bibinfo {volume} {104}},\ \bibinfo
  {pages} {115016} (\bibinfo {year} {2021}{\natexlab{b}})},\ \Eprint
  {https://arxiv.org/abs/2103.12003} {arXiv:2103.12003 [hep-ph]} \BibitemShut
  {NoStop}%
\end{thebibliography}%

\end{document}